\documentclass[preprintnumbers,showpacs,amsmath,amssymb,nofootinbib,twocolumn]{revtex4-1}
\usepackage[T1]{fontenc}
\usepackage{graphicx}
\usepackage{epsfig}
\usepackage{dcolumn}
\usepackage{multirow}
\usepackage{rotating}
\usepackage{verbatim}
\usepackage{bm}
\usepackage{color}
\usepackage{soul}
\usepackage{float}
\usepackage{hyperref}
\usepackage{comment}
\usepackage{tabu}
\usepackage[left]{lineno}
\newcommand{\ba}{\begin{array}}
\newcommand{\ea}{\end{array}}

\begin{document}
%\linenumbers

\title{Forward elastic scattering: Dynamical gluon mass and semihard interactions}

\author{M.~Broilo}
\email{mateus.broilo@ufrgs.br}
\affiliation{Instituto de F\'isica e Matem\'atica, Universidade Federal de Pelotas, 96010-900, Pelotas, RS, Brazil}
\affiliation{Instituto de F\'isica, Universidade Federal do Rio Grande do Sul, Caixa Postal 15051, 91501-970, Porto Alegre, RS, Brazil}
\author{D.~A.~Fagundes}
\email{daniel.fagundes@ufsc.br}
\affiliation{Departamento de Ci\^encias Exatas e Educa\c{c}\~ao, Universidade Federal de Santa Catarina - Campus Blumenau, 89065-300, Blumenau,
SC, Brazil}
\author{E.~G.~S.~Luna}
\email{luna@if.ufrgs.br}
\affiliation{Instituto de F\'isica, Universidade Federal do Rio Grande do Sul, Caixa Postal 15051, 91501-970, Porto Alegre, RS, Brazil}
\affiliation{Instituto de F\'{\i}sica, Facultad de Ingenier\'{\i}a, Universidad de la Rep\'ublica \\
J.H. y Reissig 565, 11000 Montevideo, Uruguay}
\author{M.~J.~Menon}
\email{menon@ifi.unicamp.br}
\affiliation{Instituto de F\'{\i}sica Gleb Wataghin, Universidade Estadual de Campinas, 13083-859, Campinas, SP, Brazil}

 \date{\today}

\begin{abstract}
The role of low-$x$ parton dynamics in dictating the high-energy behavior of forward scattering observables at LHC energies is investigated using a
QCD-based model with even-under-crossing amplitude dominance at high-energies.
We explore the effects of different sets of pre- and post-LHC fine-tuned parton distributions on the forward quantities $\sigma_{tot}$ and $\rho$, from $pp$ and $\bar{p}p$ scattering in the interval
10 GeV - 13 TeV.
We also investigate the role of the leading soft contribution, the low-energy cuttoff, and the energy dependence of the semihard form factor on these
observables.
We show that in all cases investigated the highly restrictive data on $\rho$ parameter at $\sqrt{s}=13$ TeV 
indicate that a crossing-odd component may play a crucial role in forward
elastic scattering at the highest energies, namely the Odderon
contribution.

% We show that from a statistical viewpoint 

% it is shown that the $pp$ and $\bar{p}p$ elastic
% scattering data on $\sigma_{tot}$ and $\rho$ above 10 GeV are, from a statistical viewpoint, can be described in a satisfactory way.

\end{abstract}


\maketitle

% \tableofcontents

\section{Introduction}

The elastic hadronic scattering at high energies represents a rather simple kinematic process.  However, its complete dynamical description is still a fundamental problem in QCD, since the confinement phenomena precludes a pure perturbative approach. Over the past few years, the LHC has released precise measurements of elastic proton-proton scattering which has become an important guide for selecting models and theoretical approaches, looking for a better
understanding of the theory of strong interactions.  

Among other physical observables, two \textit{forward} quantities play a fundamental role in the
investigation of the elastic scattering at high energies, the total cross section
and the $\rho$ parameter, which can be expressed in terms of the scattering amplitude
$\mathcal{A}(s,t)$ by
\begin{eqnarray}
\sigma_{tot}(s) = 4 \pi \text{Im}\,\mathcal{A}(s, t=0),
\label{sigtotrho1}
\end{eqnarray}
\begin{eqnarray}
\rho(s) = \frac{\text{Re}\,\mathcal{A}(s, t=0)}{\text{Im}\,\mathcal{A}(s, t=0)},
\label{sigtotrho2}
\end{eqnarray}
where $s$ and $t$ are the Mandelstam variables and $t=0$ indicates the forward direction.
            
Recently, the TOTEM Collaboration has provided new experimental measurements on
$\sigma_{tot}$ and $\rho$ from LHC13, the highest energy reached in accelerators.
In a first paper \cite{Antchev:2017dia}, by using as input $\rho = 0.10$, the measurement of the total cross
section yielded
\begin{eqnarray}
 \sigma_{tot}=110.6 \pm 3.4 \, \text{mb}.\nonumber
\label{datast1}
\end{eqnarray}
In a subsequent work \cite{Antchev:2017yns}, an independent measurement of the total cross section
was reported,
\begin{eqnarray}
\sigma_{tot}=110.3 \pm 3.5 \, \text{mb},\nonumber
\label{datast2}
\end{eqnarray}
together with the first measurements of the $\rho$ parameter:
\begin{eqnarray}
\rho = 0.10 \pm 0.01 \,\, \text{and} \,\,\, \rho = 0.09 \pm 0.01. \nonumber
\label{datarho}
\end{eqnarray}

Although the values of $\sigma_{tot}$ are in consensus with the increase of previous measurements
by TOTEM, the $\rho$ values indicate a rather unexpected decrease, as
compared with measurements at lower energies and predictions from the wide majority
of phenomenological models. This new information has originated a series of recent papers
and  discussions on possible phenomenological explanations for the rather small $\rho$-value. The main concern in these theoretical discussions is the full understanding of the Odderon concept (a crossing odd color-singlet
with at least three gluons) \cite{Lukaszuk:1973nt,Ewerz:2005rg,Ewerz:2003xi} and of
the Pomeron one (a crossing even color-singlet with at least two gluons) \cite{Forshaw:1997dc,Donnachie:2002en}.

The variety of recent phenomenological analyses treats different aspects involved,
pointing to distinct scenarios, and might be grouped in some classes
according to their main characteristics:

\begin{itemize}

\item
Maximal Odderon (e.g., Martynov, Nicolescu \cite{Martynov:2017zjz,Martynov:2018nyb,Martynov:2018sga,Martynov:2018pye}) and Odderon effects
in elastic hadron scattering (e.g., Cs{\"o}rg{\H o}, Pasechnik, Ster \cite{Csorgo:2018uyp,Csorgo:2019rsr}, Gon\c calves, Silva
\cite{Goncalves:2018nsp});

\item
discussions on Odderon effects in other reactions (e.g.,  Harland-Lang, Khoze, Martin, Ryskin
\cite{Harland-Lang:2018ytk}, Gon\c calves \cite{Goncalves:2018pbr});

\item
Pomeron dominance with small Odderon
contribution (e.g., Khoze, Martin, Ryskin \cite{Khoze:2018kna},
Gotsman, Levin, Potashnikova \cite{Gotsman:2018buo,Gotsman:2017ncs},
Lebiedowicz, Nachtmann, Szczurek \cite{Lebiedowicz:2018plp},
Bence, Jenkovszky, Szanyi \cite{Bence:2018ain});

\item
leading Pomeron without Odderon contribution in elastic scattering (e.g., Shabelski, Shuvaev \cite{Shabelski:2018jfq},
Broilo, Luna, Menon \cite{Broilo:2018brv, Broilo:2018els, Broilo:2018qqs}, Durand and Ha \cite{Durand:2018irx})
and in other reactions (e.g., Lebiedowicz, Nachtmann, Szczurek \cite{Lebiedowicz:2018eui});

\item
reanalyzes of the differential cross section data from TOTEM \cite{Antchev:2017yns},
indicating results for $\sigma_{tot}$ and $\rho$ at 13 TeV different from
the afore-quoted values (e.g., Pacetti, Srivastava, Pancheri \cite{Pancheri:2018yhd},
Kohara, Ferreira, Rangel \cite{Kohara:2018wng}, Cudell, Selyugin \cite{Cudell:2019mbe}).

\end{itemize}

In this rather intricate scenario, we present here a phenomenological study on the forward $pp$
and $\bar{p}p$ elastic scattering data in the region 10 GeV - 13 TeV. In our model 
the behavior of the forward quantities $\sigma_{tot}(s)$ and $\rho(s)$, given by Eqs. (\ref{sigtotrho1}) and (\ref{sigtotrho2}), are expected to be asymptotically dominated by the so-called semihard interactions. This type of process originates from hard scattering of partons which carry a very small fraction of the momenta of their parent hadrons, leading to the appearance of minijets \cite{Cline:1973kv,Pancheri:1986qg}. The latter can be viewed simply as jets with transverse energy much smaller than the total center-of-mass energy available in the hadronic collision. 
The energy dependence of the cross sections is driven mainly by semihard elementary processes that include at least one gluon in the initial state, since at low $x$ they are responsible for the dominant contribution. 

In the QCD-based formalism these partonic processes are written by means of the standard QCD cross sections convoluted with partonic distribution functions. However, these processes are potentially divergent at low transferred momenta, and for this reason they must be regularized by means of some cutoff procedure. In a nonperturbative QCD context, one natural regulator was introduced by Cornwall some time ago \cite{Cornwall:1981zr}, and since then has become an important feature in eikonalized models \cite{Luna:2005nz,Luna:2006sn,Fagundes:2011zx,Bahia:2015hha,Bahia:2015gya}. This regularization process is based on the increasing evidence that the gluon may develop a momentum-dependent mass, which introduces a natural scale able to separate the perturbative from the nonperturbative QCD region. 

Thus, taking into account the possibility that the infrared properties of QCD can, in principle, generate an effective gluon mass, we explore the nonperturbative aspects of QCD in order to describe the total cross section and the ratio of the real-to-imaginary parts of the forward elastic scattering amplitude in $pp$ and $\bar{p}p$ collisions. Most importantly,
two components are considered in our eikonal representation, one associated with the semihard
interactions and calculated from QCD and a second one associated with soft
contributions and based on the Regge-Gribov phenomenology.
Except for an odd under crossing Reggeon contribution, necessary to distinguish 
between $pp$ and $\bar{p}p$ scattering at low energies, all the dominant components
at high energies (soft and semihard) are associated with even under crossing contributions,
namely we have Pomeron dominance and absence of Odderon.

Very recently a detailed study of $\sigma_{tot}$ and $\rho$ using this formalism has been presented in \cite{Broilo:2019yuo}. This study, using the
CT14 parton distribution functions (PDFs) from a global analysis by the CTEQ-TEA group \cite{Dulat:2015mca}, has shown that, despite an overall
satisfactory description of the forward data is obtained, there is evidence that the introduction of an odd-under-crossing term in the scattering
amplitude may improve the agreement with the $\rho$ data at $\sqrt{s}=13$ TeV.
Here we have extended the previous study in several significant ways. First, we have investigated the effects of different updated sets of PDFs
beyond the CT14 \cite{Dulat:2015mca}, namely CTEQ6L \cite{Pumplin:2002vw} and MMHT \cite{Harland-Lang:2014zoa}. In this way we can compare the results obtained by using pre- and post-LHC fine-tuned PDFs.
In addition, we discuss the effects of the low-energy cuttoff, the energy 
dependence of the semihard form factor on the behavior of $\sigma_{tot}$ 
and $\rho$ and the role of the soft interaction at high energies. We also 
provide  explicit formulas and all the details related to the formalism.

The work is organized as follows. In Sect. II a short review on the concept
of the dynamical gluon mass is presented.
In Sect. III we introduce all the inputs and details concerning
our QCD-based model and in Sect. IV we specify the data set and the fit procedures.
In Sect. V the fit results are presented, followed by a discussion on the corresponding physical
interpretations and implications. Our conclusions and final remarks are the contents
of Sect. VI. Details on the
analytical parametrization for the partonic cross section is presented in appendix A.

\section{The dynamical gluon mass}
\label{sec:DGM_theory}

As pointed out in the previous section, scattering amplitudes of partons in QCD contain infrared divergences. One procedure to regulate this behavior is by means of a dynamical mass generation
mechanism which is based on the fact that the nonperturbative dynamics of QCD may generate an effective momentum-dependent mass $M_{g}(Q^{2})$ for the gluons, while
preserving the local $SU(3)_{c}$ invariance \cite{Cloet:2013jya,Aguilar:2015bud,Roberts:2016vyn}. The dynamical mass $M_{g}(Q^{2})$ introduces a natural nonperturbative scale and is linked to a finite
infrared QCD effective
charge $\bar{\alpha}_{s}(Q^{2})$. The existence of a dynamical gluon mass is strongly supported by QCD lattice results. More specifically, lattice simulations reveal that the
gluon propagator is finite in the infrared region \cite{Cucchieri:2007md,Cucchieri:2007rg,Cucchieri:2009zt,Bowman:2007du,Bogolubsky:2009dc,Oliveira:2009eh,Ayala:2012pb,Bicudo:2015rma} and this result corresponds, from the Schwinger-Dyson formalism, to a massive gluon
\cite{Cornwall:1981zr,vonSmekal:1997ohs,Aguilar:2006gr,Fischer:2006vf,Aguilar:2017dco,Fischer:2017kbq}.
It is worth mentioning that infrared-finite QCD couplings are quite usual in the literature (for a recent review, see \cite{Deur:2016tte}). In addition to the evidence
already mentioned in the lattice QCD, a finite infrared behavior of $\alpha_{s}(Q^{2})$ has been suggested, for example, in studies using QCD functional methods
\cite{Zwanziger:2001kw,Gies:2002af,Braun:2006jd}, and in studies of the Gribov-Zwanziger scenario \cite{Zwanziger:2003cf,Dudal:2008sp,Gao:2017uox}. 

Since the gluon mass generation is a purely dynamical effect, a formal continuum approach for tackling this nonperturbative
phenomenon is provided by the aforementioned Schwinger-Dyson equations that govern the dynamics of all QCD Green's functions \cite{Cornwall:1981zr,vonSmekal:1997ohs,Aguilar:2006gr,Fischer:2006vf,Aguilar:2017dco,Fischer:2017kbq,Roberts:1994dr,Alkofer:2000wg}. 
These equations constitute an infinite set of coupled nonlinear integral equations and, after a proper truncation procedure, it is possible to obtain as a
solution an infrared finite gluon propagator, while preserving the gauge invariance (or the BRST symmetry) in question. In this work we adopt 
the functional forms of $M_{g}$ and $\bar{\alpha}_{s}$ obtained by Cornwall \cite{Cornwall:1981zr} via the pinch technique in
order to derive a gauge invariant Schwinger-Dyson equation for the gluon propagator and the triple gluon vertex:
\begin{eqnarray}
M^{2}_{g}(Q^{2})=m^{2}_{g}\left[\frac{\ln\left[\left(Q^{2}+4M^{2}_{g}(Q^{2}\right)/\Lambda^{2}\right]}{\ln\left(4m^{2}_{g}/\Lambda^{2}\right)}\right]^{-12/11},
\label{dgm}
\end{eqnarray}
\begin{eqnarray}
\bar{\alpha}_{s}(Q^{2})=\frac{4\pi}{\beta_{0}\ln\left[\left(Q^{2}+4M^{2}_{g}(Q^{2})\right)/\Lambda^{2}\right]},
\label{alfaS}
\end{eqnarray}
where $\Lambda$ is the QCD scale parameter, $\beta_{0}=11-2n_{f}/3$ ($n_{f}$ is the number of flavors) and $m_{g}$ is the gluon mass scale to be phenomenologically
adjusted in order to yield well founded results in strongly interacting processes. Note that the dynamical mass $M^{2}_{g}(Q^{2})$ vanishes in the limit
$Q^{2} \gg \Lambda^{2}$. It is thus evident that in this same limit the effective charge $\bar{\alpha}_{s}(Q^{2})$ matches with the one-loop perturbative coupling:
\begin{eqnarray}
\bar{\alpha}_{s}(Q^{2}\gg \Lambda^{2})\sim \frac{4\pi}{\beta_{0}\ln(Q^{2}/\Lambda^{2})}=\alpha^{pQCD}_{s}(Q^{2}).
\label{chDGM.3}
\end{eqnarray}
In the limit $Q^{2} \to 0$, in turn, the effective charge $\bar{\alpha}_{s}(Q^{2})$ have an infrared fixed point, i.e. the dynamical mass tames the Landau pole. More
precisely, if the relation $m_{g}/\Lambda > 1/2$ is satisfied then $\bar{\alpha}_{s}(Q^{2})$ is holomorphic (analytic) on the range $0 \leq Q^{2} \leq \Lambda ^2$
\cite{Bahia:2015hha}. In fact, this is the case, since the values of the ratio $m_{g}/\Lambda$ obtained phenomenologically typically lies
in the interval $m_{g}/\Lambda \in [1.1, 2]$ \cite{Luna:2005nz,Luna:2006qp,Luna:2006sn,Luna:2010tp,Sauli:2011xr,Fagundes:2011zx,Jia:2012wf,Lipari:2013kta,Giannini:2013jla,Sidorov:2013aza,Allendes:2014fua,Beggio:2013vfa,Cvetic:2013gta,Bahia:2015hha}.

\section{QCD-based model}
\label{sec:QCD_eik}
\subsection{Eikonal Representation}

The correct calculation of high-energy hadronic interactions must be compatible with analyticity and unitarity constraints, where the latter is satisfied simply by means of eikonalized amplitudes. We adopt the following normalization for the
elastic scattering amplitude:
\begin{eqnarray}
\mathcal{A}(s,t)=i\int^{\infty}_{0}\,db\,b\,J_{0}(q_{t}b)\,\left[1-e^{-\chi(s,b)}\right],
\label{ampeik}
\end{eqnarray}
where $s$ is the square of the total center-of-mass energy, $b$ is the impact parameter, $q^{2}_{t}=-t$ is the usual Mandelstam invariant, with the complex eikonal function denoted by
\begin{eqnarray}
\chi(s,b) &=& \text{Re}\,\chi(s,b)+i\,\text{Im}\,\chi(s,b)  \nonumber \\
 &\equiv& \chi_{_{R}}(s,b)+i\,\chi_{_{I}}(s,b).
\label{eik}
\end{eqnarray}

In this picture $\Gamma(s,b)=1-e^{-\chi(s,b)}$ is the profile function, which, by the shadowing property, describes the absorption effects resulting from the opening of inelastic channels. In addition, in the impact parameter space and according to the unitarity condition of the scattering $S$-matrix it may be also written as
\begin{eqnarray}
2\text{Re}\,\Gamma(s,b)=\vert\Gamma(s,b)\vert^{2}+\left(1-e^{-2\chi_{_{R}}(s,b)}\right).
\label{unitarity}
\end{eqnarray}
Therefore, the scattering process cannot be uniquely inelastic since the elastic amplitude receives contributions from both elastic and inelastic channels. In this representation $P(s,b)=e^{-2\chi_{_{R}}(s,b)}$ can be defined as the probability that neither hadron is broken up in a collision at a given $b$  and $s$. Such an absorption factor is crucial to determine rapidity gap survival probabilities in $pp$ and $\bar{p}p$ scattering at high-energies, which in turn are crucial to disentangle inelastic diffractive (single and double) and central exclusive processes from the dominant minimum-bias (non-diffractive) cross section \cite{Fagundes:2017xli,Khoze:2017sdd}.  

Within the eikonal representation, Eq. (\ref{ampeik}), the total cross section and
the $\rho$ parameter in Eqs. (\ref{sigtotrho1}) and (\ref{sigtotrho2}) are given by:
\begin{eqnarray}
\sigma_{tot}(s)&=4\pi\int^{\infty}_{0} db\,b \left[1-e^{-\chi_{_{R}}(s,b)}\cos\chi_{_{I}}(s,b)\right];\label{eq:eikonal_sigma}\\[.5cm]
\rho(s)&= - \frac{\int^{\infty}_{0} db\,b\,e^{-\chi_{_{R}}(s,b)}\sin\chi_{_{I}}(s,b)}{\int^{\infty}_{0} db\,b\,\left(1-e^{-\chi_{_{R}}(s,b)}\cos\chi_{_{I}}(s,b)\right)}.\label{eq:eikonal_rho}
\end{eqnarray}

The eikonals for elastic $pp$ and $\bar{p}p$ scattering are connected with crossing
even ($+$) and odd ($-$) eikonals by
\begin{eqnarray}
\chi_{pp}^{\bar{p}p}(s,b) = \chi^{+} (s,b) \pm \chi^{-} (s,b).
\label{chDGM.13}
\end{eqnarray}

Real and imaginary parts of the eikonals can be connected either by Derivative Dispersion Relations (DDR) \cite{Bronzan:1974jh,Block:1984ru,Avila:2003cu,Avila:2005rg,Avila:2001qz,Avila:2002tk} or Asymptotic Uniqueness (AU), which is based on the Phragm\'en-Lindel\"off theorems  \cite{Eden:1967bk,Block:2006hy} (see \cite{Fagundes:2017iwb}, appendixes B,C,D for a recent short review on these subjects). We have tested both methods and in what follows we present the results with the AU approach, also referred to as asymptotic prescriptions or real analytic amplitudes
\cite{Block:2006hy}.

\subsection{Semihard and Soft Contributions}

The eikonal function is assumed to be the sum of the soft and the semihard (SH) parton interactions in the hadronic collision \cite{Durand:1987yv,Durand:1988ax},
\begin{eqnarray}
\chi(s,b) = \chi_{_{soft}}(s,b) + \chi_{_{SH}}(s,b),
\label{eikdef}
\end{eqnarray}
with each one related, in the general case, to the corresponding crossing even and odd contributions:
\begin{eqnarray}
\chi^{\pm}(s,b) = \chi^{\pm}_{_{soft}}(s,b) + \chi^{\pm}_{_{SH}}(s,b).
\label{chDGM.14}
\end{eqnarray}

In what follows we specify the inputs for each one of the four aforementioned  contributions
to the eikonal.

\subsubsection{Semihard Contributions and the Dynamical Gluon Mass}
\label{sec:SH_eik}

The fundamental basis of models inspired upon QCD, or also known as minijet models, is that the semihard scatterings of partons in hadrons are responsible for the observed increase of the total cross section. Here we assume a Pomeron dominance, represented by a crossing even contribution, namely we consider that the semihard odd component does not contribute with the scattering process,
\begin{eqnarray}
\chi^{-}_{_{SH}} = 0.
\nonumber
\end{eqnarray}

In respect to the even contribution, it follows from the QCD improved parton model. At leading order, this semihard eikonal can be factorized as
\begin{eqnarray}
\chi_{SH}^{\mathbf{+}}(s,b) = \frac{1}{2}\, W_{_{SH}}(s,b)\,\sigma_{QCD}(s),
\label{eiksh}
\end{eqnarray}
where $W_{_{SH}}(s,b)$ is the overlap density distribution of semihard parton scattering, $\sigma_{QCD}$ denotes the cross section of hard parton scattering in the region where pQCD can be safely applied, namely above the cutoff $Q^{2}_{min}$.

We assume (as in previous studies \cite{Bahia:2015hha}) that hard parton scattering configuration in the transverse plane of the collision (in $b$-space) to be given by the Fourier-Bessel transform:
\begin{eqnarray}
W_{_{SH}}(s,b;\nu_{\!\!_{SH}}) &=& \frac{1}{2\pi}\int_{0}^{\infty}dk_{\perp}\, k_{\perp}\, J_{0}(k_{\perp}b)\,
[G_{_{SH}}(s,k_{\perp};\nu_{_{SH}})]^{2} \nonumber \\
 &=& \frac{\nu^{2}_{_{SH}}}{96\pi} (\nu_{_{SH}} b)^{3} K_{3}(\nu_{_{SH}} b),
\label{wsh}
\end{eqnarray}
where $G_{_{SH}}(s,k_{\perp})$ is the well-known dipole parametrization
\begin{eqnarray}
G_{_{SH}}(s,k_{\perp};\nu_{_{SH}})=\left( \frac{\nu_{_{SH}}^{2}}{k_{\perp}^{2}+\nu_{_{SH}}^{2}} \right)^{2},
\label{ffsh}
\end{eqnarray}
with $\nu_{_{SH}}=\nu_{_{SH}}(s)$ taken as an energy dependent scale of the dipole. Specifically, we assume a logarithmic dependence for $\nu_{SH}$, namely:
\begin{equation}
\nu_{_{SH}}= \nu_{1}-\nu_{2}\ln (s/s_{0}),
\label{gt001}
\end{equation}
where $\nu_{1}$ and $\nu_{2}$ are two free fit parameters and the scale $\sqrt{s_{0}} = 5$ GeV is fixed. Regarding this dependence of the form factor on the energy,
though not being formally established in the context of QCD, it is truly supported by the wealth of accelerator data available (as we shall see in Section \ref{sec:fit_results}) and seems to us more realistic than taking a \textit{static} partonic configuration in $b$-space.  In addition, many other phenomenological models have been proposed in literature (see e.g. \cite{Carreras:1972fu,White:1973fr,Menon:1991ma,
Covolan:1992mf,Menon:1996es,Lipari:2009rm,Fagundes:2013aja,Fagundes:2015vba}), in which the energy dependence in form factors play a crucial role in  $pp$ and $\bar{p}p$ elastic scattering dynamics and, therefore, in accurate descriptions of the data beyond $\sqrt{s}\sim $10 GeV.  

The dynamical contribution, $\sigma_{QCD}(s)$, is calculated using perturbative QCD as follows:
\begin{eqnarray}
\sigma_{QCD}(s)&=&\sum_{ij}\,\frac{1}{1+\delta_{ij}}\,\int^{1}_{0} dx_{1}\int^{1}_{0} dx_{2}\int^{\infty}_{Q^{2}_{min}}\!\!\!\!d\vert\hat{t}\vert\,\frac{d\hat{\sigma}_{ij}}{d\vert\hat{t}\vert}(\hat{s},\hat{t})\nonumber\\
& \times & f_{i/A}(x_{1},\vert\hat{t}\vert)\,f_{j/B}(x_{2},\vert\hat{t}\vert)\,\Theta\left(\frac{\hat{s}}{2}-\vert \hat{t} \vert\right),
\label{sigQCD}
\end{eqnarray}
where $x_{1}$ and $x_{2}$ are momentum fraction carried by partons in the hadrons $A$ and $B$, respectively,
$\hat{s}=x_{1}x_{2}s$, $\vert\hat{t}\vert\equiv Q^{2}$ stands for Mandelstam invariants of parton-parton scatterings such as e.g. $gg\rightarrow gg$, $qg\rightarrow qg$ and $gg\rightarrow \bar{q}q$ (whose partonic cross sections are given afterwards) and $f_{i/A}(x_1, |\hat{t}|)$, $f_{j/B}(x_2, |\hat{t}|)$ are the parton distribution functions (PDFs) for partons i and j. The indexes $i,j=q,\bar{q},g$ identify quark (anti-quark) and gluon degrees of freedom and $Q^{2}_{min}$ represent the minimum momentum transfer scale allowing for pQCD calculations of partonic hard scattering, obeying the constraint $2Q^{2}_{min}<2\vert \hat{t}\vert<\hat{s}$. 

Concerning the differential cross section at elementary level, the major contribution at high energies are the ones initiated by gluons\footnote{Despite the potential influence of \textit{soft} gluon radiation at the initial state, such as discussed in \cite{Fagundes:2015vba} and references therein, we only consider the effects of gluon radiation in the Parton Distribution Functions, as following from DGLAP evolution.}

\begin{itemize}

\item[i.] gluon-gluon elastic scattering,
\begin{eqnarray}
\frac{d\hat{\sigma}}{d\hat{t}}(gg\to gg)=\frac{9\pi\bar{\alpha}^{2}_{s}}{2\hat{s}^{2}}\left(3 -\frac{\hat{t}\hat{u}}{\hat{s}^{2}}-
\frac{\hat{s}\hat{u}}{\hat{t}^{2}}-\frac{\hat{t}\hat{s}}{\hat{u}^{2}} \right) ,
\label{ggst}
\end{eqnarray}

\item[ii.] quark-gluon elastic scattering,
\begin{eqnarray}
\frac{d\hat{\sigma}}{d\hat{t}}(qg\to qg)=\frac{\pi\bar{\alpha}^{2}_{s}}{\hat{s}^{2}}\, (\hat{s}^{2}+\hat{u}^{2}) \left(
\frac{1}{\hat{t}^{2}}-\frac{4}{9\hat{s}\hat{u}} \right) ,
\label{qgst}
\end{eqnarray}

\item[iii.] gluon fusion into a quark pair,
\begin{eqnarray}
\frac{d\hat{\sigma}}{d\hat{t}}(gg\to \bar{q}q)=\frac{3\pi\bar{\alpha}^{2}_{s}}{8\hat{s}^{2}}\, (\hat{t}^{2}+\hat{u}^{2}) \left(
\frac{4}{9\hat{t}\hat{u}}-\frac{1}{\hat{s}^{2}} \right),
\label{chDGM.37}
\end{eqnarray}
\end{itemize}
 with kinematical constraints imposed and  \textit{connected with the dynamical mass}, namely: (i) $\hat{s}+\hat{t}+\hat{u}=4M^{2}_{g}(Q^{2})$, for gluon elastic scattering ($gg\rightarrow gg$) and (ii)  $\hat{s}+\hat{t}+\hat{u}=2M^{2}_{g}(Q^{2})+2M^{2}_{q}(Q^{2})$ for gluon fusion ($gg\rightarrow \bar{q}q$) and quark-gluon scattering $qg\rightarrow qg$. Importantly,  in what follows we assume the Cornwall's dynamical gluon mass (in Euclidean space) \cite{Cornwall:1981zr}, Eq. (\ref{dgm}), with the infrared frozen effective QCD charge, Eq. (\ref{alfaS}), to interpolate two QCD domains:
(i) $Q^{2}\approx 0$, i.e. at infrared, where $M^{2}_{g}$ freezes and the gluons carries an effective bare mass, $M^{2}_{g}(0)=m^{2}_{g}$; (ii) $Q^{2} \gg m^{2}_{g}, \Lambda^{2}$, dynamical mass generation from nontrivial vacuum structure becomes unimportant and perturbative QCD limit is achieved. 
 
As discussed in Section \ref{sec:DGM_theory}, recent phenomenology and lattice studies support bare gluon masses in the range, $m_{g}:300-700$ MeV. Here we fix 
\begin{eqnarray}
m_g = 400 \, \mathrm{MeV} \nonumber
\end{eqnarray}
while also accounting, for completeness, the subdominant role of dynamical quark generation at high energies. We assume
\begin{eqnarray}
M_{q}(Q^{2})=\frac{m^{3}_{q}}{Q^{2}+m^{2}_{q}},
\label{eq:Mq_parametrization}
\end{eqnarray}
which also recovers the bare mass $m_{q}$ (with $m_{q}<m_{g}$) at infrared and reaches the massless quark limit for $Q^{2}\gg m^{2}_{q}$. In all calculations we take
\begin{eqnarray}
m_q = 250 \, \mathrm{MeV}
\nonumber
\end{eqnarray}
as fixed scale. At last, as commented before, the \textit{complex} eikonal $\chi_{SH}^{+}(s,b)$ is determined through the asymptotic even prescription $s \rightarrow -i s$. The parametrization for $\sigma_{QCD}(s)$, Eq. (18), and the real 
and imaginary parts provided by the above prescription are presented 
and discussed in Appendix A. We notice that in case of $W_{_{SH}}(s,b;\nu_{\!\!_{SH}})$, Eq. (15),
the prescription results in a complex Bessel function of complex
argument.

% The details on this dependence and the evaluation of the real and imaginary parts of $\sigma_{QCD}(s)$ are presented and discussed in Appendix A.

\subsubsection{Soft Contributions}

The full even and odd soft contributions are based on the Regge-Gribov formalism
and are constructed in accordance with Asymptotic Uniqueness (Phragm\'en-Lindel\"off theorems).
Assuming also leading even component, they are parametrized by
\begin{eqnarray}
\chi_{_{soft}}^{+}(s,b) = \frac{1}{2}\,W^{+}_{_{soft}}(b;\mu^{+}_{soft})\sigma^{+}(s),
\label{chDGM.10}
\end{eqnarray}
\begin{eqnarray}
\chi_{_{soft}}^{-}(s,b) = \frac{1}{2}\, W^{-}_{_{soft}}(b;\mu^{-}_{_{soft}})\sigma^{-}(s),
\label{chDGM.32}
\end{eqnarray}
where  
\begin{eqnarray}
\sigma^{+}(s) = A+\frac{B}{\sqrt{s/s_{0}}}e^{i\pi/4}+C \left[ \left(s/s_{0}\right) e^{-i\pi /2} \right]^{\lambda} , 
\label{chDGM.33}
\end{eqnarray}
\begin{eqnarray}
\sigma^{-}(s) = D\, \frac{e^{-i\pi/4}}{\sqrt{s/s_{0}}},
\end{eqnarray}
denote analytical even and odd cross sections. Here $\lambda = 0.12$ and $A$, $B$, $C$ and $D$ are free fit parameters.
Moreover, the impact parameter structure derives from bidimensional Fourier transform of dipole form factors, namely:
\begin{eqnarray}
W^{+}_{_{soft}}(b;\mu^{+}_{_{soft}}) &=& \frac{1}{2\pi}\int_{0}^{\infty}dk_{\perp}\, k_{\perp}\, J_{0}(k_{\perp}b)\,G_{dip}^{2}(k_{\perp};\mu^{+}_{_{soft}}) \nonumber \\
 &=& \frac{(\mu^{+}_{_{soft}})^{2}}{96\pi} (\mu^{+}_{_{soft}} b)^{3} K_{3}(\mu^{+}_{_{soft}} b),
\label{chDGM.19}
\end{eqnarray}
\begin{eqnarray}
W^{-}_{_{soft}}(b;\mu^{-}_{_{soft}}) = \frac{(\mu^{-}_{_{soft}})^{2}}{96\pi} (\mu^{-}_{_{soft}} b)^{3} K_{3}(\mu^{-}_{_{soft}} b),
\label{chDGM.20}
\end{eqnarray}
where $\mu^{-}_{_{soft}} \equiv 0.5$ GeV is a fixed parameter and $\mu^{+}_{_{soft}}$ a free fit parameter. As in the case of the SH form factor, the energy scale is fixed at $\sqrt{s_0} = 5$ GeV.

We notice that in the Regge-Gribov context, the soft even contribution consists of a Regge pole with intercept $\alpha_{\mathbb{R}}^{+}(0)=1/2$, a critical Pomeron and a single-pole Pomeron, with intercept $\alpha_{\mathbb{P}}(0)=1+\lambda$.
The odd contribution is associated with only a Regge pole, with intercept $\alpha_{\mathbb{R}}^{-}(0)=1/2$. 

Summarizing the model has 7 free fit parameters, 5 associated with the soft contribution, $A, B, C, D, \mu^{+}_{_{soft}}$ and only 2 with the semihard contribution, $\nu_1$ and  $\nu_2$ (from $\nu_{SH}(s)$ in $W^{+}_{_{SH}}(s,b)$). In addition, 5 parameters are fixed: $m_g = 400$ GeV, $m_q = 250$ GeV, $s_0 = 25$ GeV$^{2}$, $\mu^{-}_{soft} = 0.5$ GeV and $\lambda = 0.12$.

\section{Dataset and Fit Procedures}
\label{sec:fit_results}

In the absence of ab initio theoretical QCD arguments to determine the parameters $A, B, C, D, \mu^{+}_{_{soft}}$, $\nu_1$ and  $\nu_2$, we resort to a fine-tuning fit procedure described in what follows. As we are interested in the very high-energy behavior of $\sigma_{tot}$ and $\rho$, we shall use only $pp$ and $\bar{p}p$ elastic scattering data.
Moreover, in order to test our QCD-based model in the $t=0$ limit, we perform global fits that include exclusively forward data, given by Eqs. (\ref{eq:eikonal_sigma}) and (\ref{eq:eikonal_rho}). 

\subsection{Dataset}

Our dataset is compiled from a wealth of collider data on $pp$ and $\bar{p}p$ elastic scattering, available in the Particle Data Group (PDG) database \cite{Tanabashi:2018oca} as well as in the very recent papers of LHC Collaborations such as TOTEM \cite{Antchev:2016vpy,Antchev:2017yns,Antchev:2017dia,Antchev:2018rec} and ATLAS \cite{Aaboud:2016ijx,Aad:2014dca}, which span a large c.m. energy range, namely 10 GeV $\leqslant \sqrt{s} \leqslant 13 $ TeV. For the sake of clarity and completeness we furnish in Table \ref{tab:dataLHC_notPDG} all the recent LHC data on
$\sigma_{tot}$ and $\rho$, still absent in the PDG2018 review. This dataset totalizes 13 new data points on $pp$ forward elastic scattering at high energies.

We call attention to the fact that we do not apply to this dataset, composed of 174 data points on $\sigma_{tot}^{\bar{p}p,pp}$ and $\rho^{\bar{p}p,pp}$, any sort of selection or sieving procedure, which might introduce bias in the analysis.

\begin{table}[h!]
\centering
\caption{Total cross section, $\sigma_{tot}$, and $\rho$-parameter data recently measured by TOTEM and ATLAS Collaborations at the LHC,  but not compiled in the PDG2018 review \cite{Tanabashi:2018oca}.}
\vspace*{.2cm}
\begin{tabular}{c@{\quad}|c@{\quad}|c@{\quad}|c@{\quad}|c@{\quad}}
\hline \hline 
& & & & \\[-0.3cm]
$\sqrt{s}$ (TeV) & $\sigma_{\text{tot}}$[mb] & $\rho$ & Collaboration & Ref. \\ 
\hline 
& & & & \\[-0.3cm]
\multirow{3}{*}{13} & $110.6 \pm 3.4$ & $-$ & TOTEM  & \cite{Antchev:2017dia}\\
 & \multirow{2}{*}{$110.3 \pm 3.5$} & $0.10\pm 0.01$ &  \multirow{2}{*}{TOTEM} &\multirow{2}{*}{\cite{Antchev:2017yns}} \\
&  & $0.09\pm 0.01$ &  & \\ 
\hline
& & & & \\[-0.3cm]
\multirow{6}{*}{8.0}  & $-$ & $0.12\pm 0.03$ & \multirow{3}{*}{TOTEM} &\multirow{3}{*}{\cite{Antchev:2016vpy}}\\
& $102.9\pm 2.3$ & $-$ &  & \\
  & $103.0\pm 2.3$ & $-$ &  & \\
 & $96.07\pm 0.92$ & $-$ & ATLAS & \cite{Aaboud:2016ijx}\\  
  & $101.5\pm 2.1$ & $-$ & \multirow{2}{*}{TOTEM} & \multirow{2}{*}{\cite{Antchev:2015zza}}\\  
  & $101.9\pm 2.1$ & $-$ &  & \\   
\hline
& & & & \\[-0.3cm]
\multirow{2}{*}{7.0}  & $99.1\pm 4.3$  & $-$ & TOTEM &  \cite{Antchev:2013iaa}\\ 
  & $95.35\pm1.36 $  & $-$ & ATLAS & \cite{Aad:2014dca} \\
\hline
& & & & \\[-0.3cm]
\multirow{1}{*}{2.76}  & $84.7\pm 3.3$  & $-$ & TOTEM & \cite{Nemes:2018tk} \\ 
\hline \hline 
\end{tabular} 
\label{tab:dataLHC_notPDG}
\end{table}

\subsection{Fit Procedures}
To provide statistical information on fit quality, we perform a best-fit analysis, furnishing as goodness of fit parameters the chi-squared per degrees of freedom ($\chi^{2}/\zeta$) and the corresponding integrated probability, $P(\chi^{2},\zeta)$ \cite{Bev:2002st}. Since our model is highly nonlinear, numerical data reduction is called for. Despite the limitation of treating statistical and systematical uncertainties at the same foot, we apply the $\chi^{2}/\zeta$ tests to our dataset with uncertainties summed in quadrature\footnote{For very recent applications of the frequentist and Bayesian approaches to high-energy elastic scattering data analysis see Refs. \cite{Orava:2018ddo,Selyugin:2018uob}}. Our fits are done using the TMINUIT class of the ROOT framework \cite{Brun:1997pa}, through the MIGRAD algorithm. While the number of calls of the MIGRAD routine may vary in the fits with PDFs CETQ6L, CT14 and MMHT, 
full convergence of the algorithm was always achieved. Moreover, all data reductions were performed with the interval $\chi^{2}-\chi^{2}_{min}=8.18$, which corresponds to 68.3 $\%$ of Confidence Level ($1\sigma$) \cite{James:1975dr} in our case (7 free parameters).

Furthermore, in all fits performed we set the low energy cutoff, $\sqrt{s_{min}}=10$ GeV.
%To test the predictive power of the model we set three possible high-energy cutoffs, namely: $\sqrt{s_{max}}=13,\ 8$ and $7$ TeV. Such method aims at testing possible influence of high-energy data such as those recently released by TOTEM in getting accurate description of data at and beyond LHC13. 
\begin{figure}[h!]
\centering
\includegraphics*[width=8.5cm,height=8cm]{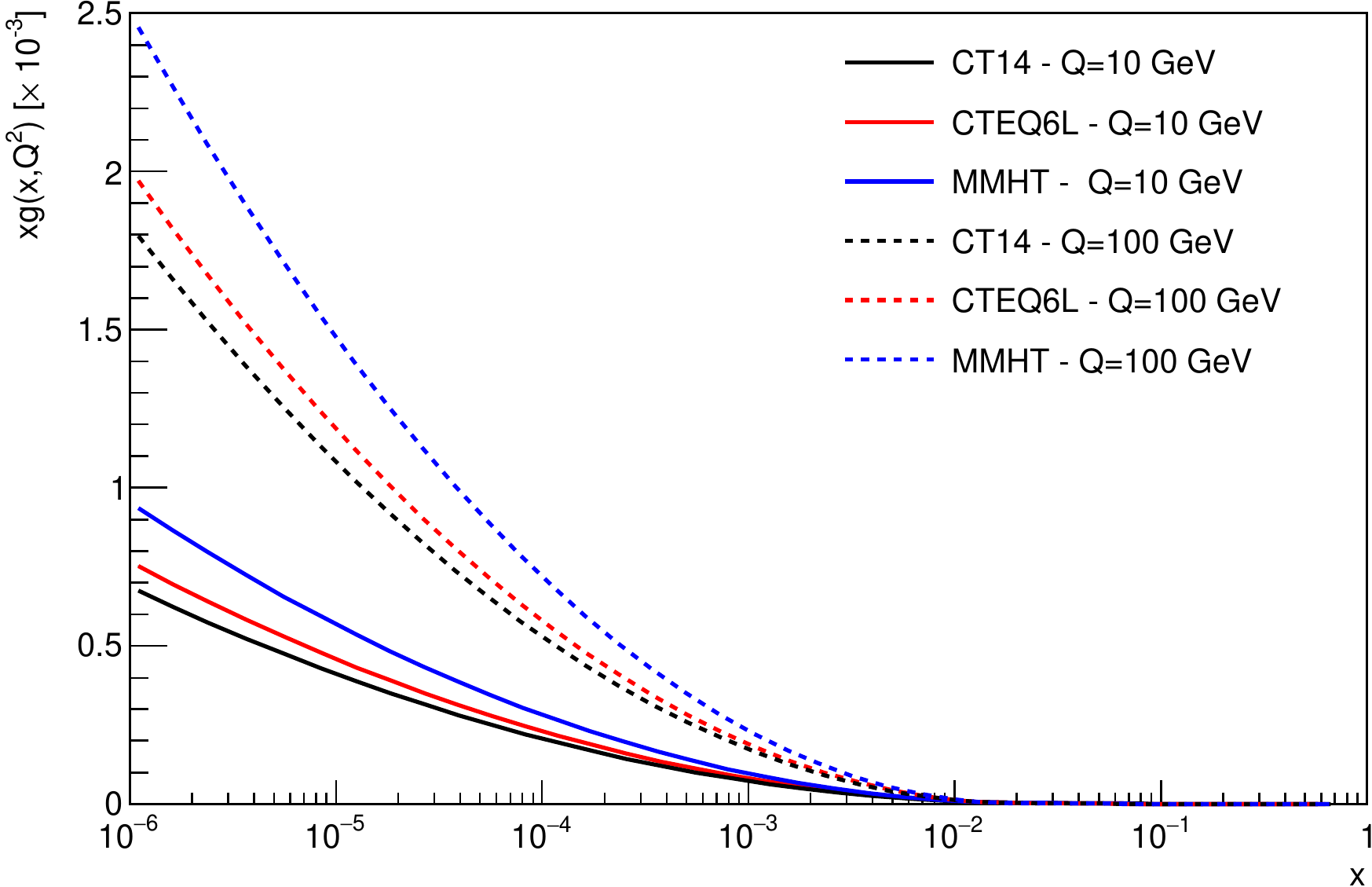}
\caption{Gluon distribution function, $xg(x,Q^{2})$, following from DGLAP evolution for PDFs, CT14, CTEQ6L and MMHT at $Q=10$ GeV and $Q=100$ GeV.}
\label{fig:gluonPDFs}
\end{figure}

\begin{figure}[h!]
\centering
\includegraphics*[width=8.5cm,height=8cm]{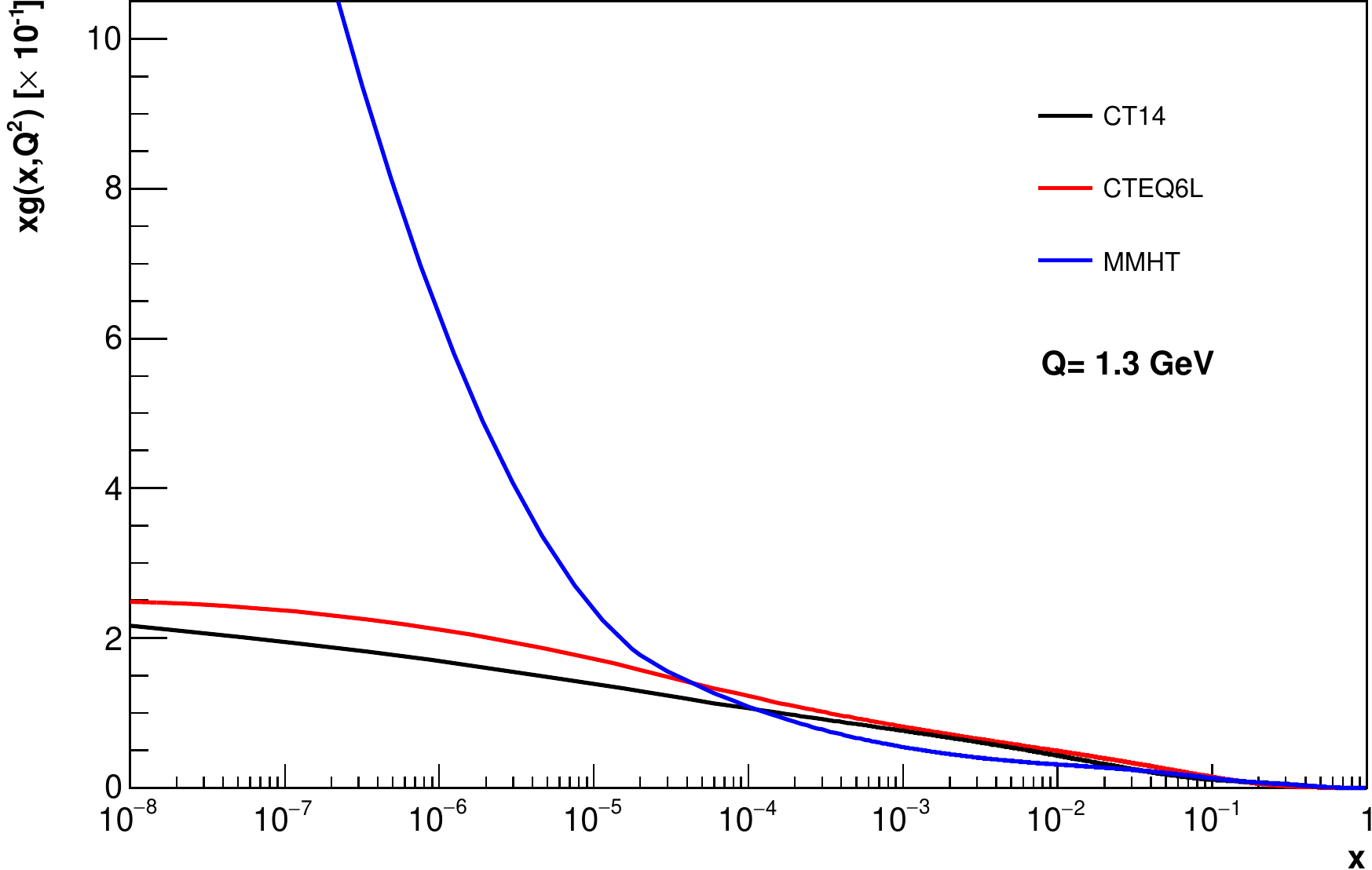}
\caption{The same as Fig. 1, but for the scale $Q = 1.3$ GeV.}
\label{fig2:gluonPDFs}
\end{figure}

In the following we present our results, according to the choice of three distinct PDFs: CTEQ6L \cite{Pumplin:2002vw} (pre-LHC), CT14 \cite{Dulat:2015mca} and MMHT \cite{Harland-Lang:2014zoa} (fine-tuned with LHC data). In testing different PDFs we look for a better understanding of the impact of low-$x$ parton dynamics in defining the very high-energy behavior of  $\sigma^{\bar{p}p,pp}_{tot}$ and $\rho^{\bar{p}p,pp}$.
For comparison and further discussion, the behavior of the gluon distribution function in each PDF set is given in Figs. \ref{fig:gluonPDFs} and
\ref{fig2:gluonPDFs}.

\section{Results and Discussion}

\begin{figure}[h!]
\centering
\includegraphics*[width=8.5cm,height=8cm]{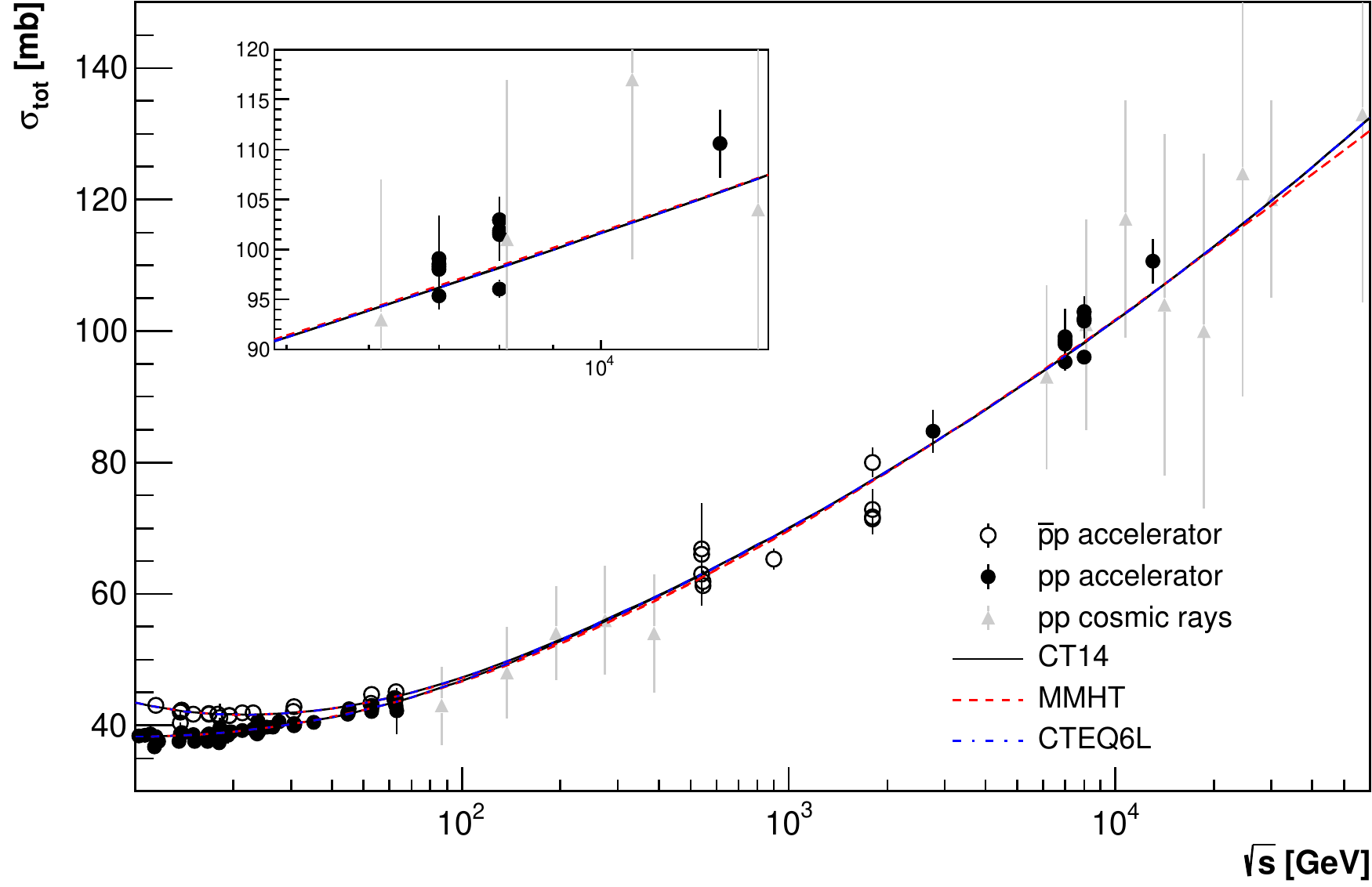}
\includegraphics*[width=8.5cm,height=8cm]{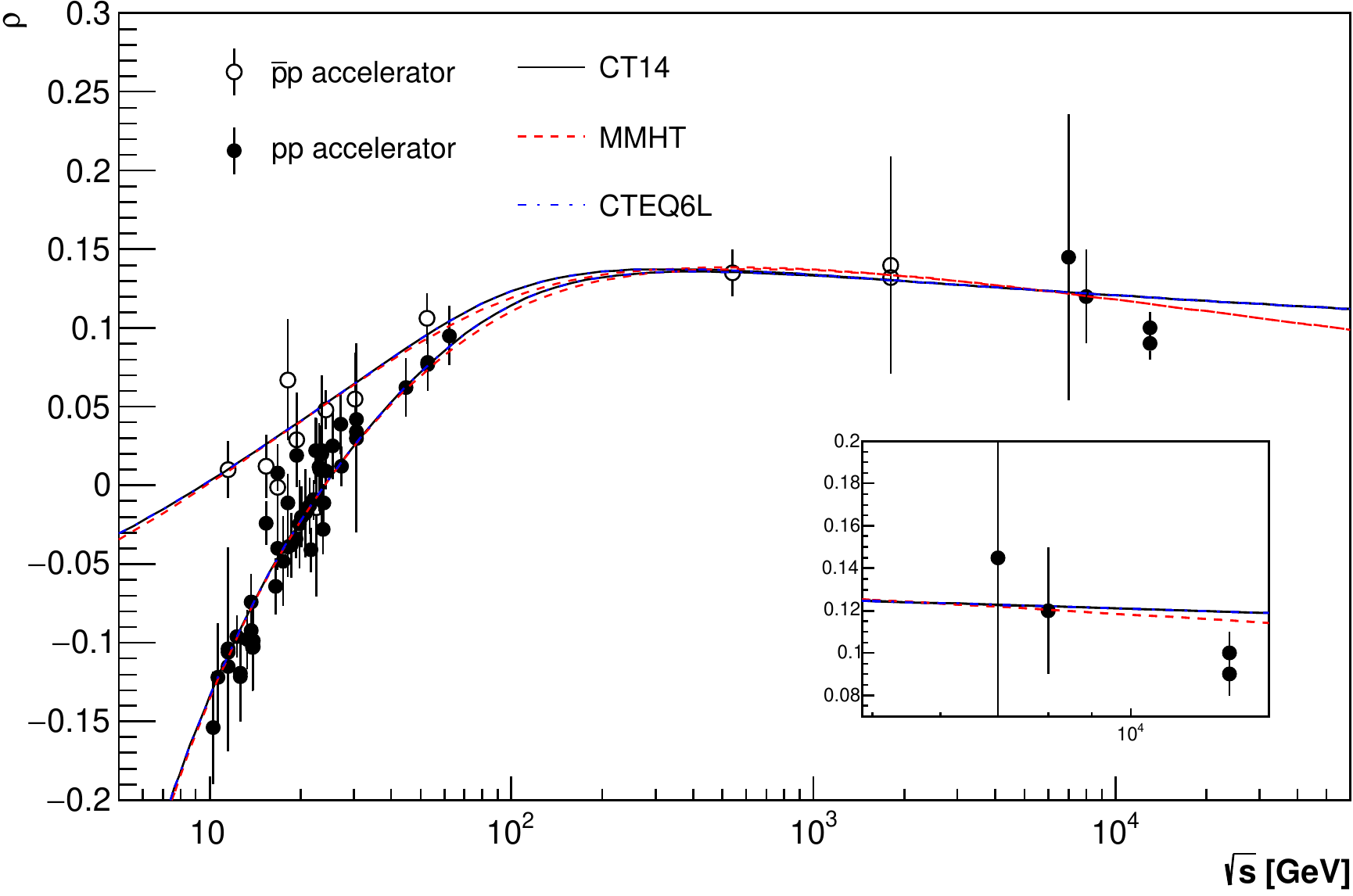}
\caption{Global 1$\sigma$-fit of total cross section, $\sigma_{tot}^{pp/\bar{p}p}$ and $\rho^{pp/\bar{p}p}$ parameter.
Best fit parameters and quality estimators are given in Table \ref{t:paramsDGM}.}
\label{fig:DGM_cut13}
\end{figure}

The results for the free fit parameters, using each one of the three PDFs (CTEQ6L, CT14, MMHT) are displayed in Table \ref{t:paramsDGM}, together with the
statistical information on the data reductions (reduced chi square and corresponding integrated probability). The curves of $\sigma_{tot}(s)$ and $\rho(s)$
for the three PDFs, compared with the experimental data, are shown in Figure \ref{fig:DGM_cut13}. Predictions of $\sigma_{tot}$ and $\rho$ for $pp$ scattering at some energies of interest are displayed
in Table \ref{tab:predictions}.

\begin{table}[h!]
\caption{Best fit parameters of the QCD-based model with PDFs CTEQ6L \cite{Pumplin:2002vw}, CT14 \cite{Dulat:2015mca} and MMHT \cite{Harland-Lang:2014zoa}.
Quality fit estimators, chi-squared per degree of freedom, $\chi^2/ \zeta$, and integrated probability, $P(\chi^2;\zeta)$, are also furnished (where $\zeta=167$ specifies the number of degrees of freedom (dof) in each fit).}
\centering
%\scalebox{0.9}{
\begin{tabular}{c@{\quad}c@{\quad}c@{\quad}c@{\quad}}
\hline \hline
& & &  \\[-0.3cm]
PDF:    & CTEQ6L & CT14 & MMHT \\[0.05ex]
\hline
& & &   \\[-0.3cm]
$\mu^{+}_{soft}$ [GeV]& 0.90\,$\pm$\,0.18  & 0.90\,$\pm$\,0.18 & 0.90\,$\pm$\,0.20 \\[0.15cm]
$A$ [GeV$^{-2}$]      & 101\,$\pm$\,11  & 88.1\,$\pm$\,9.7   & 93\,$\pm$\,11  \\[0.15cm]
$B$ [GeV$^{-2}$]      & 48\,$\pm$\,11     & 51.7\,$\pm$\,9.9      & 54\,$\pm$\,10     \\[0.15cm]
$C$ [GeV$^{-2}$]      & 16.0\,$\pm$\,6.8       & 27.3 \,$\pm$\,5.8   & 19.6\,$\pm$\,6.7           \\[0.15cm]
$\mu^{-}_{soft}$ [GeV]& 0.5 (fixed)          & 0.5 (fixed)           & 0.5 (fixed)          \\[0.15cm]
$D$ [GeV$^{-2}$]      & 24.2\,$\pm$\,1.4     & 24.2\,$\pm$\,1.4     & 24.2\,$\pm$\,1.4   \\[0.15cm]
$\nu_{1}$ [GeV]       & 1.63\,$\pm$\,0.20    & 1.70\,$\pm$\,0.22     & 1.46\,$\pm$\,0.21    \\[0.15cm]
$\nu_{2}$ [GeV]   & 0.009\,$\pm$\,0.013    & 0.015\,$\pm$\,0.014   & -0.007\,$\pm$\,0.013 \\[0.05ex]
& & &   \\[-0.3cm]
\hline
& & &   \\[-0.3cm]
$\chi^2/\zeta$   &   1.285            &      1.304          &      1.259                   \\[0.15cm]
$P(\chi^2;\zeta)$    & 7.6 $\times$ 10$^{-3}$ &  5.0 $\times$ 10$^{-3}$ & 1.3 $\times$ 10$^{-2}$ \\[0.05ex]
& & &   \\[-0.3cm]
\hline \hline 
\end{tabular}
%}
\label{t:paramsDGM}
\end{table} 

\begin{table}[H]
\centering
\caption{Predictions of our model using CT14 as a representative case.}
\vspace*{.3cm}
\begin{tabular}{c|c|c}
\hline\hline
$\sqrt{s}$ [TeV] & $\sigma_{tot}$ [mb] & $ \rho $ \\ 
\hline 
0.9 & 68.75 & 0.1338 \\ 
\hline 
2.76 & 82.91 & 0.1279 \\ 
\hline 
13 & 105.8 & 0.1194 \\ 
\hline 
14 & 107.0 & 0.1190 \\ 
\hline\hline
\end{tabular} 
\label{tab:predictions}
\end{table}

From Figure \ref{fig:DGM_cut13}  we see that, although the model provides a quite good description of the forward data in the interval 
10 GeV - 8 TeV, the results at 13 TeV do not reach the error bars of the TOTEM data on $\sigma_{tot}$
and $\rho$.

Before proceeding with further tests, let us discuss
some physical aspects related to our results.

By showing the values in the Table \ref{t:paramsDGM}, we can see that the parameter $\mu^{+}_{soft}$ has, in general, the value 0.90 GeV. This restriction is due to the fact that the
inverse of both $\mu^{+}_{soft}$ and $\mu^{-}_{soft}$ parameters characterizes the range of these soft interactions. Since the odd soft eikonal $\chi_{soft}^{-}(s,b)$ is
more sensitive to the longer-range trajectories, $\rho$ and $\omega$ exchanges, it is expected the inverse of the odd exchanges, $(\mu^{-}_{soft})^{-1}$,
to be larger than the inverse of the even ($a_{2}$ and $f_{2}$) exchanges, $(\mu^{+}_{soft})^{-1}$. Thus in our analysis we impose the reasonable condition
$1 < \mu^{+}_{soft}/\mu^{-}_{soft} \leq 1.8$. Indeed, in all cases the parameter $\mu^{+}_{soft}$ fall within the expected range.
 
In what concerns high-energy QCD dynamics,
in QCD-based (s-channel) models like ours, the driving mechanism behind the rapid rise of the total cross section is linked to the growth with energy of low-$p_{t}$ jets (called minijets). This idea, while proposed many years ago, remains a powerful one in the scope of models of strong interactions at high-energies, as it provides a clear connection between perturbative QCD and hadronic elastic observables, such as $\sigma_{tot}$ and $\rho$, in a unitarized framework. 

Those minijets arise from partonic interactions (mainly gluons) carring very small momentum fraction of their parent hadrons. On the one hand, from eq. (\ref{sigQCD}), we see that the smallest $x$ scale probed by this model is
\begin{equation}
x_{min}=\frac{Q^{2}_{min}}{s},\nonumber
\end{equation}
which, taking $Q^{2}_{min} \simeq 1$ GeV$^{2}$, yields $x_{min} \sim 10^{-10}$ at LHC13. On the other, it is well-known that at very low-x the PDF's diverge, as gluon emissions - which naturally occur in any partonic process at high energies - are not suppressed by DGLAP evolution at higher momentum transferred.  This behavior can be readily seen from Figure \ref{fig:gluonPDFs} and \ref{fig2:gluonPDFs} where the gluon distribution function from parton distributions CT14, CTEQ6L and MMHT are displayed at the minimum scale $Q_{min}=1.3$ GeV and two higher scales, $Q =10$ GeV and 100 GeV. From these plots one may notice that MMHT grows faster than CT14 and CTEQ6L, specially at low momentum scales, such as $Q_{min}=1.3$ GeV. 

As matter of fact, very low-x gluons are the key ingredient to understand our results for various PDF's, as shown in Figure   
\ref{fig:DGM_cut13}. Once the QCD cross section (\ref{sigQCD}) is dominated by low-x partons, and gluon iniciated processes are the leading component of this cross section, one expects the magnitude of $\sigma_{SH}(s)$ calculated with MMHT to be larger than the corresponding curves for CT14 and CTEQ6L at high energies. As we show in Figure \ref{fig:sig_QCD} in Appendix A, that turns out to be exactly the case.

In addition, looking for some insights into the formalism, it may be important to
notice the effects of two phenomenological inputs, one related to
the soft even eikonal and the other to the semihard form factor.
In the first case, $\chi_{soft}^{+}(s,b)$ as given by Eq. (\ref{chDGM.10}),
has a component which increases with the energy, namely the term
with coefficient $C$. In the second case, the dipole form factor 
$G_{_{SH}}(s,k_{\perp};\nu_{_{SH}})$, Eqs. (\ref{ffsh}) and (\ref{gt001}),
also depends on the energy through the logarithmic.
The effect of these terms can be investigated by assuming either $C=0$ or
$\nu_2=0$ and re-fitting the dataset.
Moreover, the efficience of the model for different
choices of the datasets is also important to be checked.
All these three variants are pressented and discussed in the
following three subsections.

\subsection{Effect of the leading contribution in $\chi^{+}_{soft}(s,b)$}

The soft-even component of the eikonal, Eqs. (\ref{chDGM.10}) and (\ref{chDGM.33}), 
comprise a leading Pomeron contribution given by the power term in Eq. (\ref{chDGM.33}), with coefficient $C$.  In order to investigate the relevance of this leading soft contribution at high energies in our global results, we present here a test in which this term is excluded. Specifically, we fix $C = 0$ in Eq. (\ref{chDGM.33}) and refit the dataset.  The results of these fits are presented in Table \ref{tab:Ctest} and Fig. 4.

% Let us compare the results in Fig. 3 ($C$ free fit parameter) with those in Fig. 4 ($C=0$ fixed), focusing the TOTEM data at 13 TeV (inserts) in the cases of PDFs CT14 and CTEQ6L.
% From Fig. 3, the results for $\sigma_{tot}$ cross the middle of the lower error bar and for $\rho$ they cross the central value of the highest measurement. On the other hand, from Fig. 7 the results for $\sigma_{tot}$ barely reach the end of the lower error bar and for $\rho$ they cross the middle of the upper error bar.

In respect the statistical quality of the fits, comparison of Tables \ref{t:paramsDGM} ($C$ free parameter)
and \ref{tab:Ctest} ($C=0$ fixed) shows that the exclusion of this contribution results in a rather unacceptable
goodness of fit, since $\chi^2/\zeta$ increase to 1.3 - 1.8 and $P(\chi^2)$ decrease at least one order
of magnitude.
For example, in case of CT14, from Table \ref{t:paramsDGM} ($C$ free parameter), $\chi^2/\zeta$ = 1.304, $P(\chi^2)$ = 5.0 $\times$ 10$^{-3}$
and from Table \ref{tab:Ctest} ($C=0$ fixed), $\chi^2/\zeta$ = 1.836, $P(\chi^2)$ = 2.4 $\times$ 10$^{- 10}$.

We conclude that, although not being the leading contribution at the highest energies,
the single pole Pomeron in the soft component is important for an adequate fit result in statistical grounds. Moreover, the faster decrease of $\rho$ observed at LHC energies can related to the correlations among low-energy parameters such as $A$ and $B$ and high-energy ones, as $\nu_{1}$ and $\nu_{2}$. In addition to the lower statistical significance of this fits, in comparison with the previous ones, the reduction of $B$ central values by a factor one-half and  the large uncertainties in $\nu_{2}$ seems to corroborate this hypothesis.

\begin{table}[h!]
\caption{Best-fit parameters of our model, obtained by fixing $C=0$, with $\zeta$ = 168 degrees of freedom.}
\centering
%\scalebox{0.9}{
\begin{tabular}{c@{\quad}c@{\quad}c@{\quad}c@{\quad}}
\hline \hline
& & &  \\[-0.3cm]
PDF:    & CTEQ6L & CT14 & MMHT \\[0.05ex]
\hline
& & &   \\[-0.3cm]
$\mu^{+}_{soft}$ [GeV]& 0.70\,$\pm$\,0.16  & 0.700\,$\pm$\,0.015   & 0.700\,$\pm$\,0.034  \\[0.15cm]
$A$ [GeV$^{-2}$]      & 111.31\,$\pm$\,0.70  & 117.86\,$\pm$\,0.62   & 109.9\,$\pm$\,0.77  \\[0.15cm]
$B$ [GeV$^{-2}$]      & 28.0\,$\pm$\,3.0     & 12.1\,$\pm$\,2.8      & 28.0\,$\pm$\,3.1     \\[0.15cm]
$C$ [GeV$^{-2}$]      & 0 (fixed)            & 0 (fixed)             & 0 (fixed)            \\[0.15cm]
$\mu^{-}_{soft}$ [GeV]& 0.5 (fixed)          & 0.5 (fixed)           & 0.5 (fixed)          \\[0.15cm]
$D$ [GeV$^{-2}$]      & 23.4\,$\pm$\,1.3     & 23.6\,$\pm$\,1.3      & 23.5\,$\pm$\,1.3     \\[0.15cm]
$\nu_{1}$ [GeV]       & 1.68\,$\pm$\,0.12    & 1.54\,$\pm$\,0.13     & 1.41\,$\pm$\,0.12    \\[0.15cm]
$\nu_{2}$ [GeV]   & 0.012\,$\pm$\,0.0084      & 0.0072\,$\pm$\,0.0091   & -0.010\,$\pm$\,0.0087 \\[0.05ex]
& & &   \\[-0.3cm]
\hline
& & &   \\[-0.3cm]
$\chi^2/\zeta$   & 1.338               & 1.836               & 1.385                        \\[0.15cm]
$P(\chi^2;\zeta)$    & 2.3 $\times$ 10$^{-3}$ & 2.4 $\times$ 10$^{-10}$ & 7.1 $\times$ 10$^{-4}$ \\[0.05ex]
& & &   \\[-0.3cm]
\hline \hline 
\end{tabular}
%}
\label{tab:Ctest}
\end{table}

\begin{figure}[h!]
\centering
\includegraphics*[width=8.5cm,height=8cm]{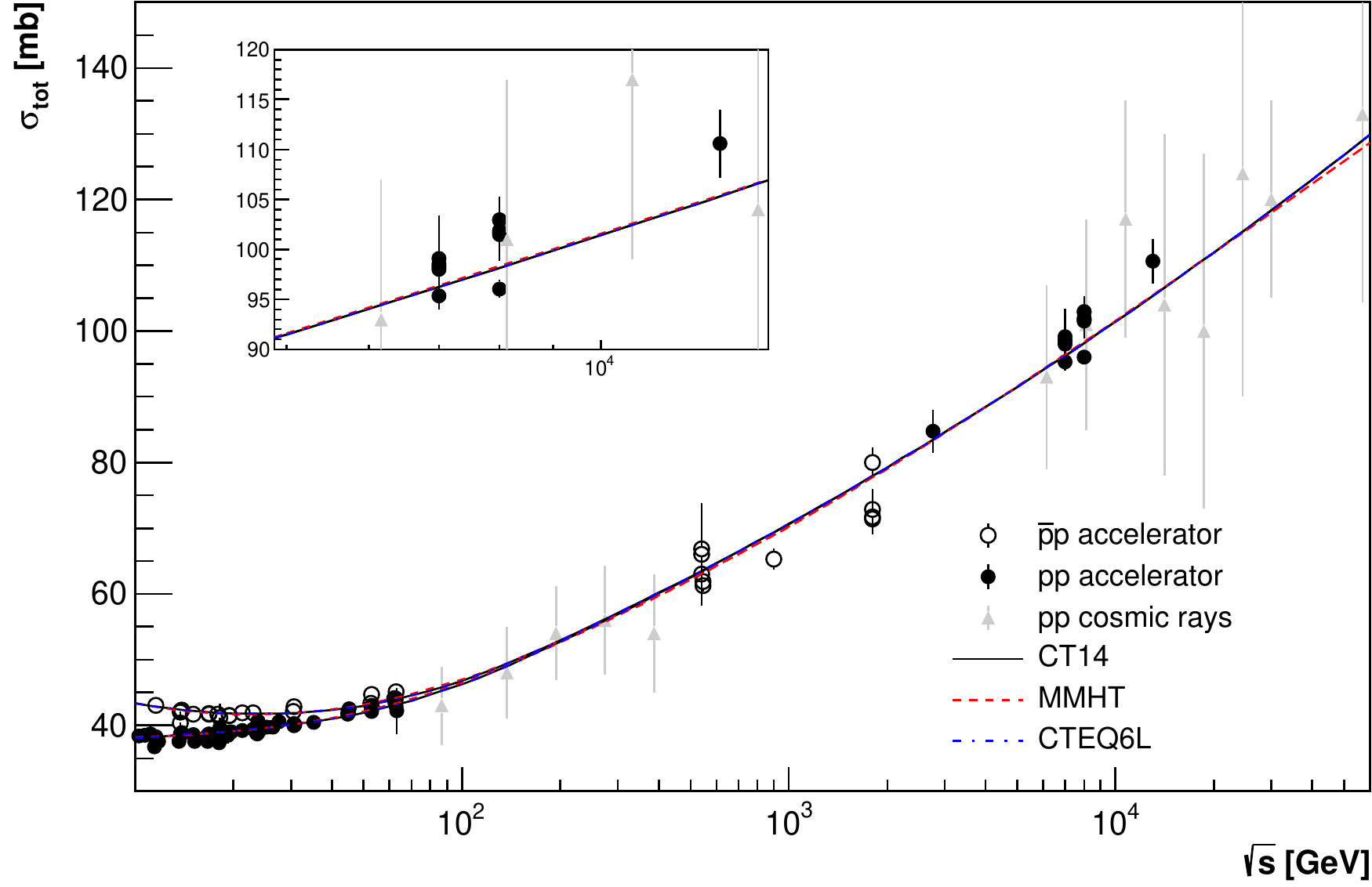}
\includegraphics*[width=8.5cm,height=8cm]{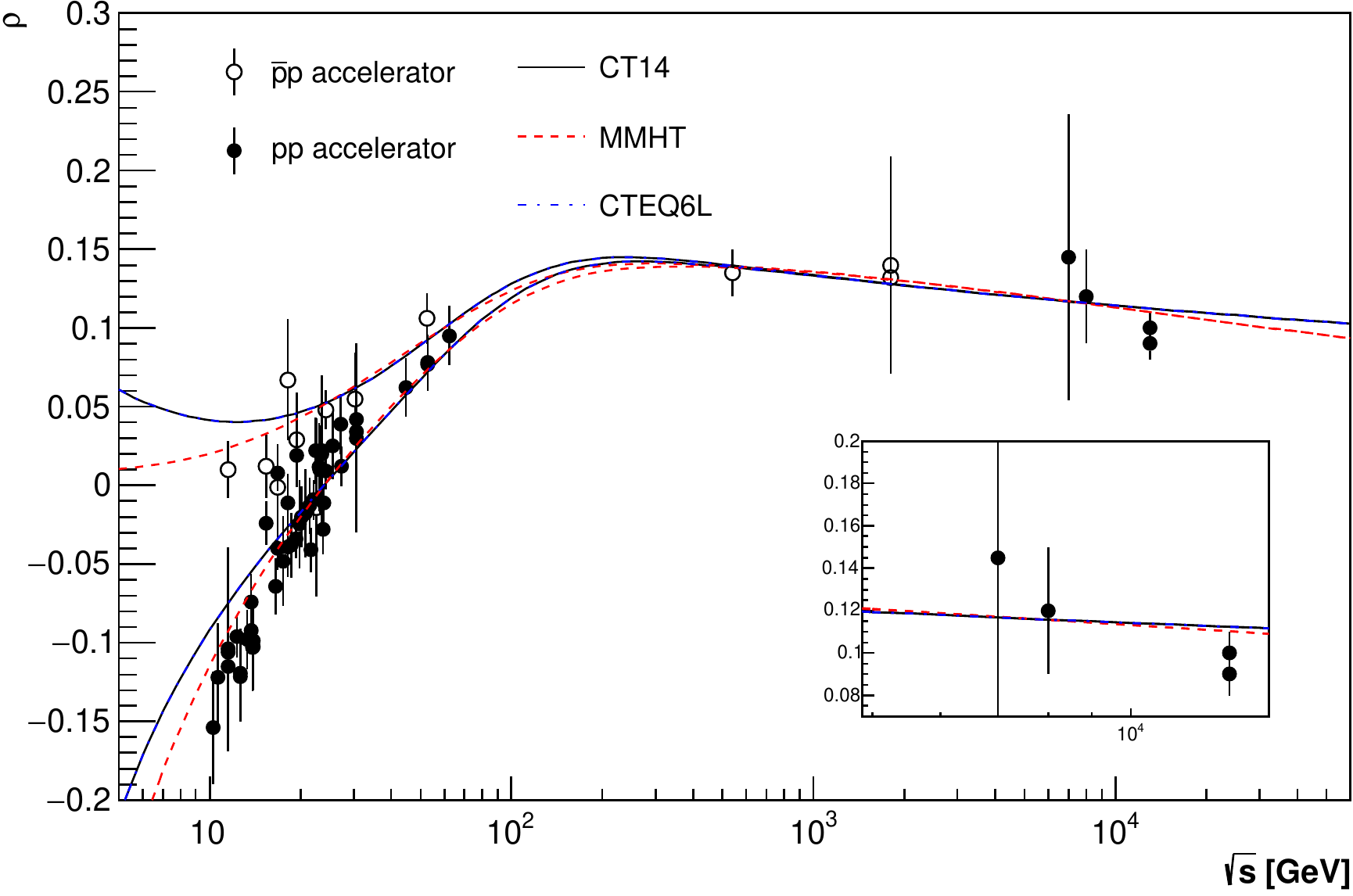}
\caption{Global 1$\sigma$-fit to $\sigma_{tot}^{pp/\bar{p}p}$ and $\rho^{pp/\bar{p}p}$ data without rising soft terms in $\chi^{+}_{soft}(s,b)$.  Statistical information is provided in Table \ref{tab:Ctest}.}
\label{fig:TestC}
\end{figure}

\subsection{Effect of the energy dependence in the semihard form factor}

\begin{figure}[h!]
\centering
\includegraphics*[width=8.5cm,height=8cm]{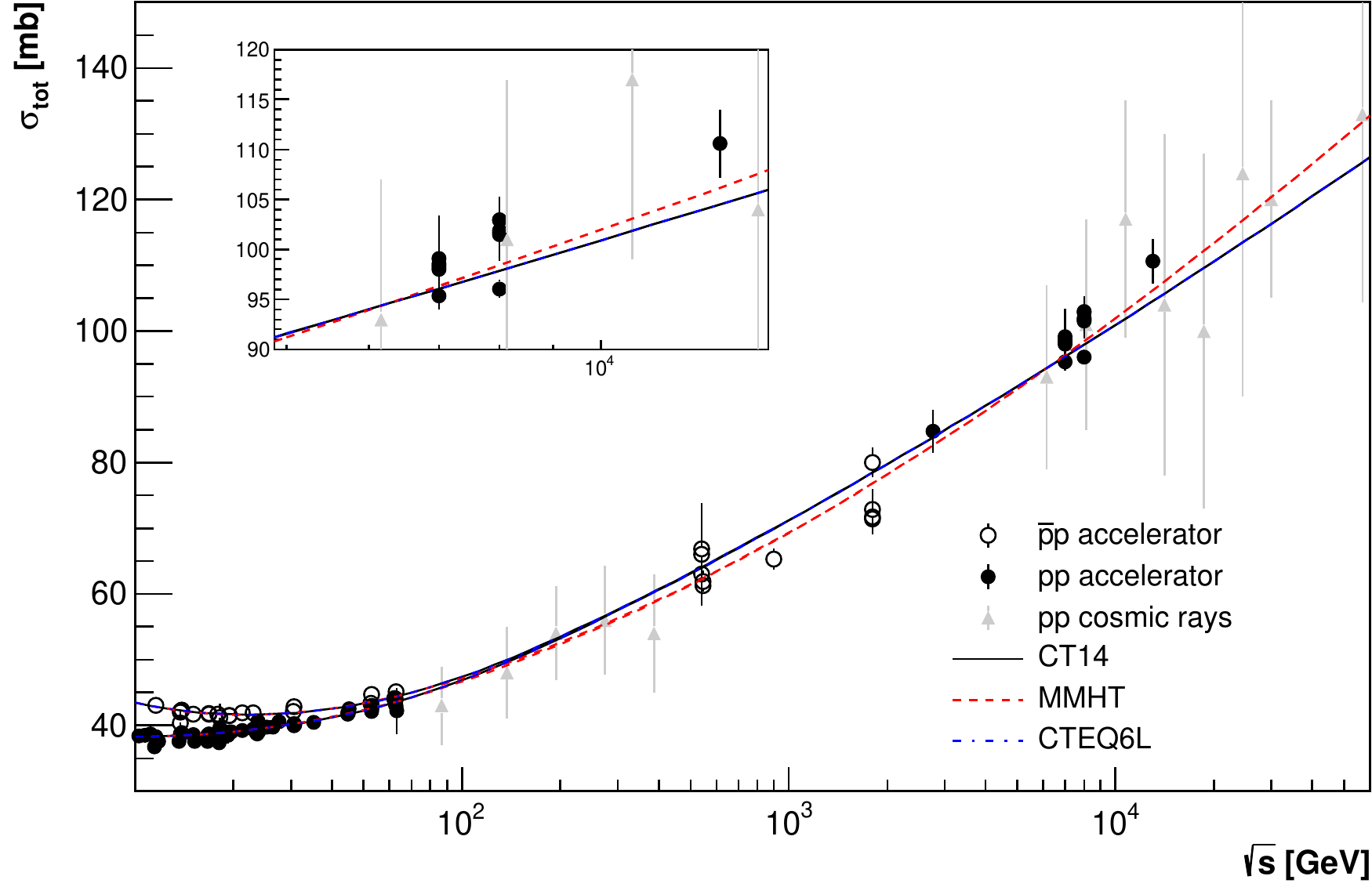}
\includegraphics*[width=8.5cm,height=8cm]{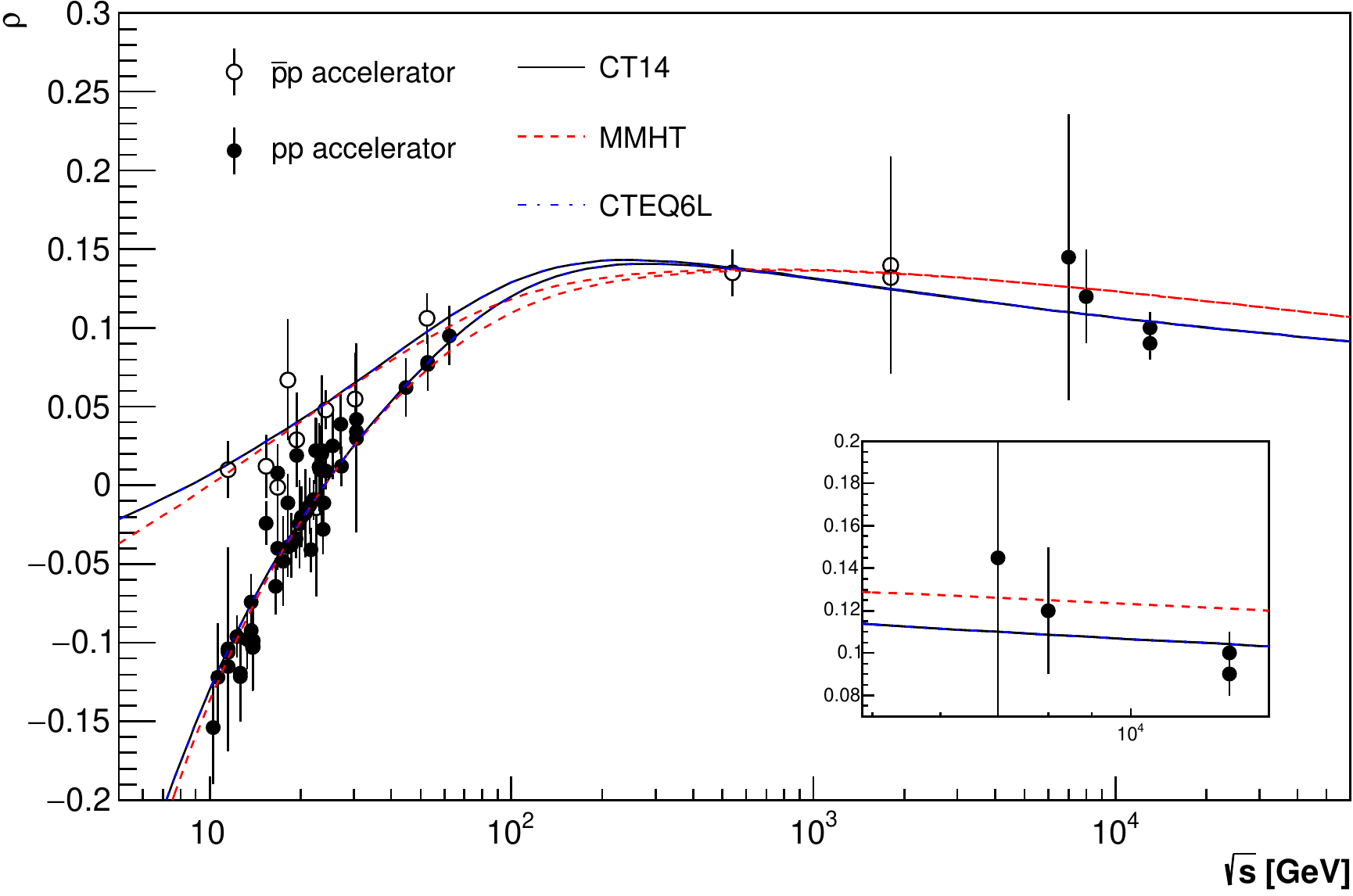}
\caption{Global 1$\sigma$-fit to $\sigma_{tot}^{pp/\bar{p}p}$ and $\rho^{pp/\bar{p}p}$ data taking $\nu_{2}$=0. Best fit parameters and quality estimators are given in Table \ref{tab:nu2test}}
\label{fig:nu2test}
\end{figure}

Although not so usual in the present phenomenological context, 
one of the ingredients of the QCD-based model is the energy dependence embodied in the
semihard form factor, Eqs. (\ref{wsh}) and (\ref{ffsh}). As commented in our introduction, this assumption is associated with
the possibility of a broadening of the spacial gluon distribution as the energy increases.
In order to investigate the relevance of this assumption in our global results, we present
here a test in which this energy dependence is excluded. Specifically, we fix $\nu_2 = 0$ in Eq. (\ref{gt001}),
so that $\nu_{SH} = \nu_{1}$ and
refit the data set. As before, we consider 
the three PDFs employed in this work.
The results of these fits are presented in Table \ref{tab:nu2test} and Fig. \ref{fig:nu2test}.

Comparison of Tables II ($\nu_2$ free) and V ($\nu_2$ = 0),
shows that the statistical quality of the fits ($\chi^2/ \zeta$, 
$P(\chi^2;\zeta)$) are similar, but with a slight increase (decrease) in
$\chi^2/ \zeta$ ($P(\chi^2;\zeta)$) in case of $\nu_2$ = 0. From Figures \ref{fig:DGM_cut13} ($\nu_2$ free)
and \ref{fig:nu2test} ($\nu_2$ = 0), we notice distinct behaviors related to the
use of the MMHT on one side and CT14/CTEQ6 on the other.
This may be related with the faster rise of $\sigma_{QCD}(s)$ in
case of MMHT, as compared with the slower rise within
CT14 and CTEQ6L (see Fig. 9 in Appendix A).

We also notice that at 13 TeV, MMHT leads to the highest values of both
$\sigma_{tot}$ and $\rho$. For CT14 and CTEQ6L, although the $\rho$
results reach the upper error bar, those for $\sigma_{tot}$
lie far below the lower error bar.

\begin{table}[h!]
%\scalebox{0.9}{
\centering
\caption{Best-fit parameters of our model, obtained by fixing $\nu_{2}=0$, with $\zeta$ = 168 degrees of freedom.}
\begin{tabular}{c@{\quad}c@{\quad}c@{\quad}c@{\quad}}
\hline \hline
& & &  \\[-0.3cm]
PDF:    & CTEQ6L & CT14 & MMHT \\[0.05ex]
\hline
& & &   \\[-0.3cm]
$\mu^{+}_{soft}$ [GeV]& 0.90\,$\pm$\,0.15    & 0.90\,$\pm$\,0.13     & 0.84\,$\pm$\,0.13    \\[0.15cm]
$A$ [GeV$^{-2}$]      & 106.0\,$\pm$\,8.1    & 93.7\,$\pm$\,7.8     & 92\,$\pm$\,14    \\[0.15cm]
$B$ [GeV$^{-2}$]      & 44.6\,$\pm$\,8.7     & 47.6\,$\pm$\,8.5      & 51\,$\pm$\,30     \\[0.15cm]
$C$ [GeV$^{-2}$]      & 12.8\,$\pm$\,4.7    & 23.7\,$\pm$\,4.5   & 17\,$\pm$\,24
\\[0.15cm]
$\mu^{-}_{soft}$ [GeV]& 0.5 (fixed)          & 0.5 (fixed)           & 0.5 (fixed)          \\[0.15cm]
$D$ [GeV$^{-2}$]      & 24.2\,$\pm$\,1.3     & 24.2\,$\pm$\,1.3      & 24.0\,$\pm$\,1.8     \\[0.15cm]
$\nu_{1}$ [GeV]       & 1.486\,$\pm$\,0.031  & 1.469\,$\pm$\,0.034   & 1.588\,$\pm$\,0.078  \\[0.15cm]
$\nu_{2}$ [GeV]       & 0 (fixed)            & 0 (fixed)             & 0 (fixed)            \\[0.15cm]
& & &   \\[-0.3cm]
\hline
& & &   \\[-0.3cm]
$\chi^2/\zeta$   & 1.304               & 1.362               & 1.263                        \\[0.15cm]
$P(\chi^2;\zeta)$    & 4.9 $\times$ 10$^{-3}$ & 1.3 $\times$ 10$^{-3}$ & 1.2 $\times$ 10$^{-2}$ \\[0.05ex]
& & &   \\[-0.3cm]
\hline \hline 
\end{tabular}
%}
\label{tab:nu2test}
\end{table} 

\vspace*{-0.8cm}

\subsection{Changing the data-set}

Here we develop two tests on the efficiency of the QCD-based model related to two different choices
of the dataset. In the first test the low-energy cutoff is lowered from 10 GeV down to 5 GeV and in the second
test the ATLAS data at 7 and 8 TeV are not included in the dataset.
We present the results obtained with the three PDFs.
Since the results are similar to those presented in the main text with our standard dataset,
we focus the discussion on those obtained
with the PDF CT14.

\subsubsection{Low-energy cutoff down to 5 GeV}

\begin{figure}[h!]
\centering
\includegraphics*[width=8.5cm,height=8cm]{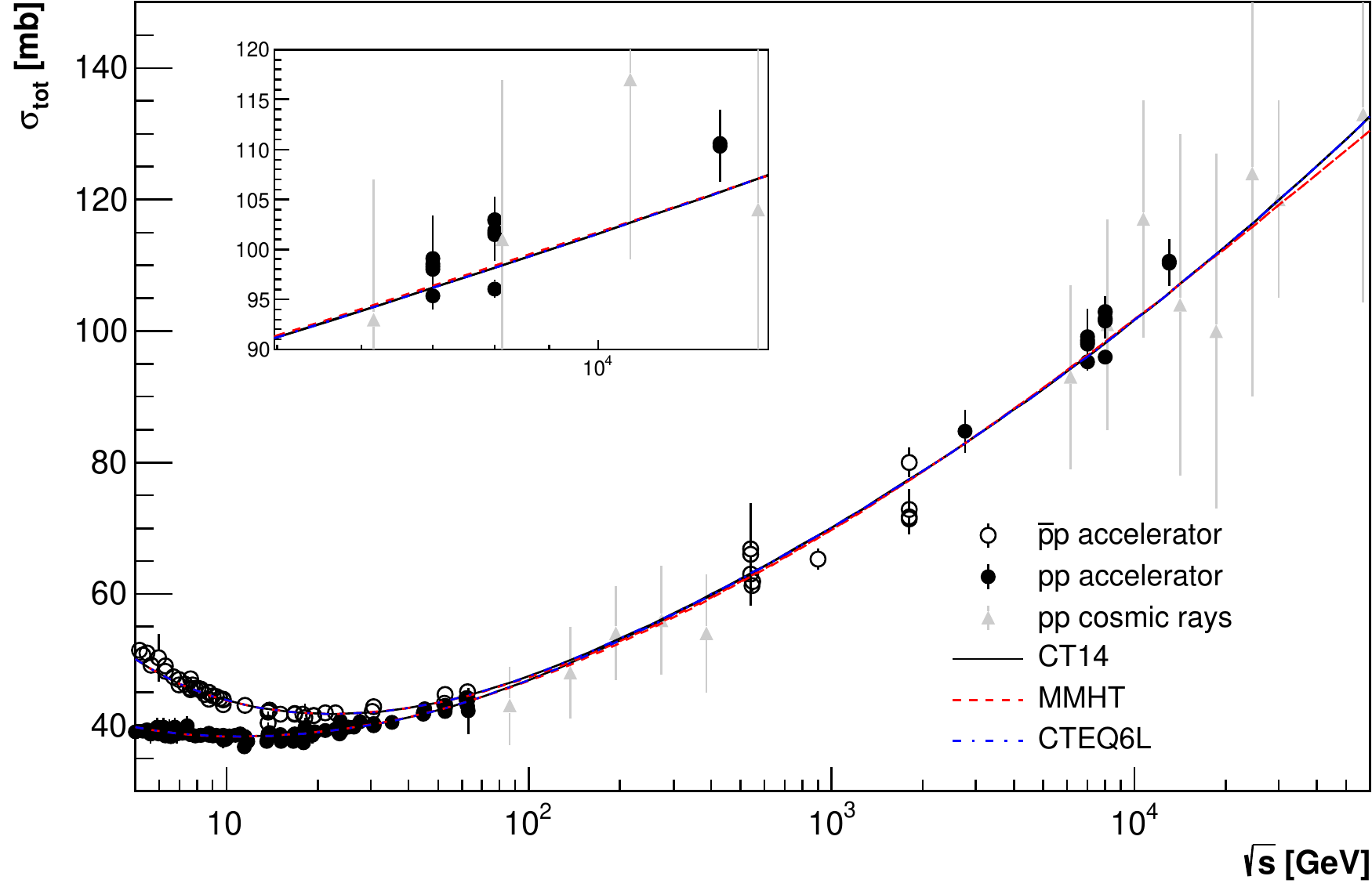}
\includegraphics*[width=8.5cm,height=8cm]{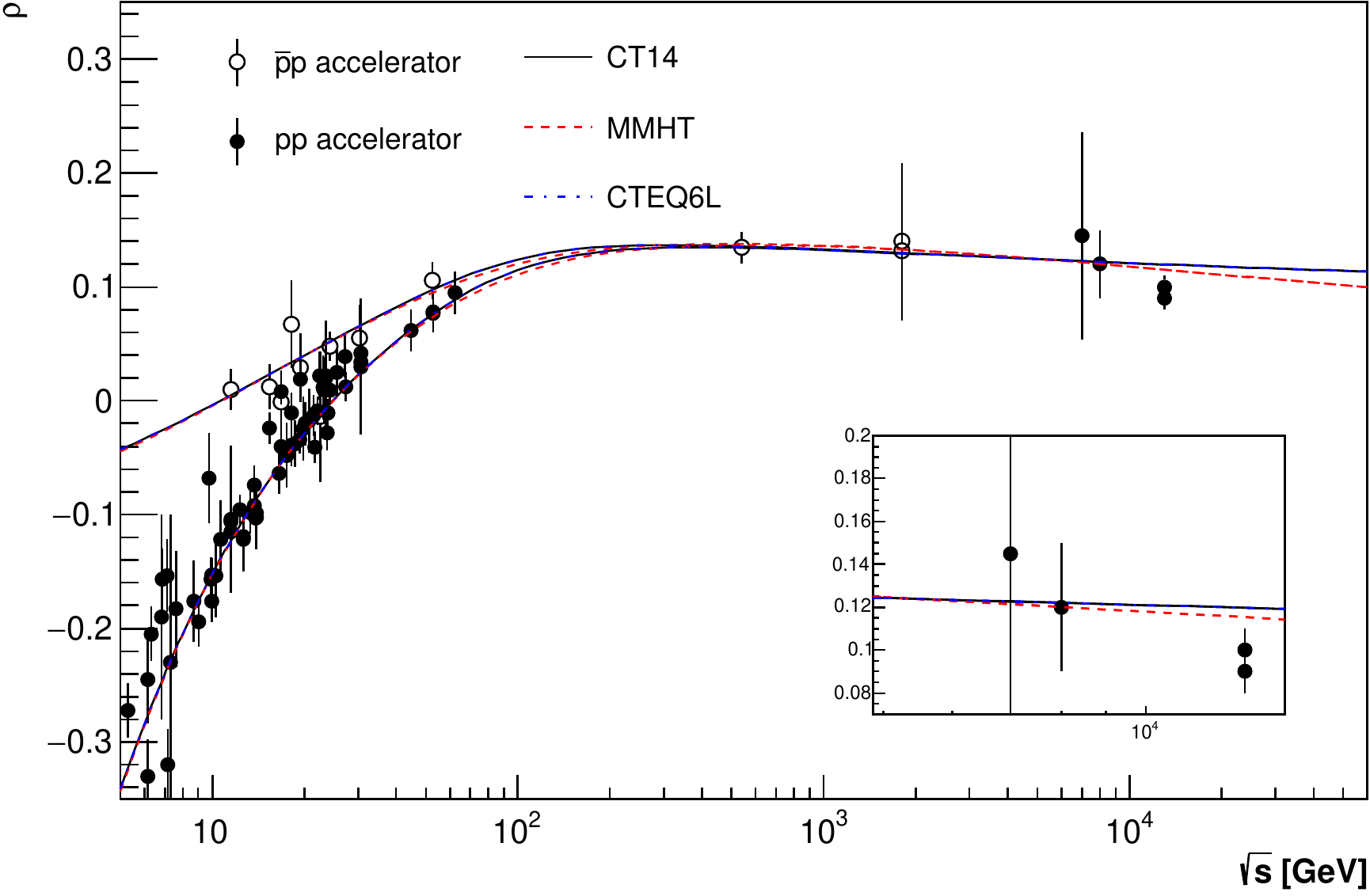}
\caption{Global 1$\sigma$-fit to $\sigma_{tot}^{pp/\bar{p}p}$ and $\rho^{pp/\bar{p}p}$ data for low-energy cutoff $\sqrt{s_{min}}=5$ GeV.}
\label{fig:ATLAS_cuts5-13}
\end{figure}

By lowering the energy cutoff to 5 GeV, we add 85 points for $\sigma_{tot}$
and $\rho$ in the dataset. The result of the fit is displayed in Table \ref{tab:smin5GeV_comATLAS} and Fig. \ref{fig:ATLAS_cuts5-13}, indicating, within CT14,
$\chi^2/\zeta$ = 1.464, for $\zeta$ = 251 and $P(\chi^2;\zeta)$  = $2.2\,\times\,10^{-6}$.
Our results with cutoff at 10 GeV are shown in Fig. \ref{fig:DGM_cut13} and Table \ref{t:paramsDGM} (CT14) and in this case, $\chi^2/\zeta$ = 1.304, for $\zeta$ = 167 and $P(\chi^2;\zeta)$  = $5.0\,\times\,10^{-3}$.

Although the integrated probability decreases three orders of magnitude for $\sqrt{s_{min}}$ = 5.0 GeV, from Figures \ref{fig:DGM_cut13} and \ref{fig:ATLAS_cuts5-13}, we see that we see that the visual description of the data is quite good and the quality of the fit is reasonable for this data set (without any sieve procedure), showing that the model
can cover efficiently the whole region 5 GeV - 8 TeV. Still, the problem of simultaneously fitting $\sigma_{tot}$ and $\rho$ at 13 TeV (within uncertainties) remains.

\begin{table}[h!]
%\scalebox{0.9}{
\centering
\caption{Best-fit parameters of our model with low-energy cut-off  $\sqrt{s_{min}}$ = 5.0 GeV. The number of degrees of freedom is $\zeta$ = 251.}
\vspace*{.2cm}
\begin{tabular}{c@{\quad}c@{\quad}c@{\quad}c@{\quad}}
\hline \hline
& & &  \\[-0.3cm]
PDF:    & CTEQ6L & CT14 & MMHT \\[0.05ex]
\hline
& & &   \\[-0.3cm]
$\mu^{+}_{soft}$ [GeV]& 0.90\,$\pm$\,0.16    & 0.90\,$\pm$\,0.16     & 0.90\,$\pm$\,0.17    \\[0.15cm]
$A$ [GeV$^{-2}$]      & 92.5\,$\pm$\,2.9   & 82.1\,$\pm$\,5.9    & 88.1\,$\pm$\,6.9    \\[0.15cm]
$B$ [GeV$^{-2}$]      & 58.3\,$\pm$\,4.5     & 60.0\,$\pm$\,4.1      & 62.1\,$\pm$\,4.4     \\[0.15cm]
$C$ [GeV$^{-2}$]      & 21.1\,$\pm$\,4.6    & 30.7\,$\pm$\,3.7 	  & 22.8\,$\pm$\,4.7    \\[0.15cm]
$\mu^{-}_{soft}$ [GeV]& 0.5 (fixed)          & 0.5 (fixed)           & 0.5 (fixed)          \\[0.15cm]
$D$ [GeV$^{-2}$]      & 26.03\,$\pm$\,0.74   & 26.02\,$\pm$\,0.74    & 26.02\,$\pm$\,0.74   \\[0.15cm]
$\nu_{1}$ [GeV]       & 1.70\,$\pm$\,0.20    & 1.76\,$\pm$\,0.21     & 1.49\,$\pm$\,0.21   \\[0.15cm]
$\nu_{2}$ [GeV]       & 0.012\,$\pm$\,0.013  & 0.018\,$\pm$\,0.014   & -0.006\,$\pm$\,0.014 \\[0.15cm]
& & &   \\[-0.3cm]
\hline
& & &   \\[-0.3cm]
$\chi^2/\zeta$   & 1.451                  & 1.464                 & 1.430                    \\[0.15cm]
$P(\chi^2;\zeta)$    & 3.8 $\times$ 10$^{-6}$ & 2.2 $\times$ 10$^{-6}$ & 8.8 $\times$ 10$^{-6}$   \\[0.05ex]
& & &   \\[-0.3cm]
\hline \hline 
\label{tab:smin5GeV_comATLAS}
\end{tabular}
\end{table}

\subsubsection{Fits without the ATLAS data}

\begin{figure}[h!]
\centering
\includegraphics*[width=8.5cm,height=8cm]{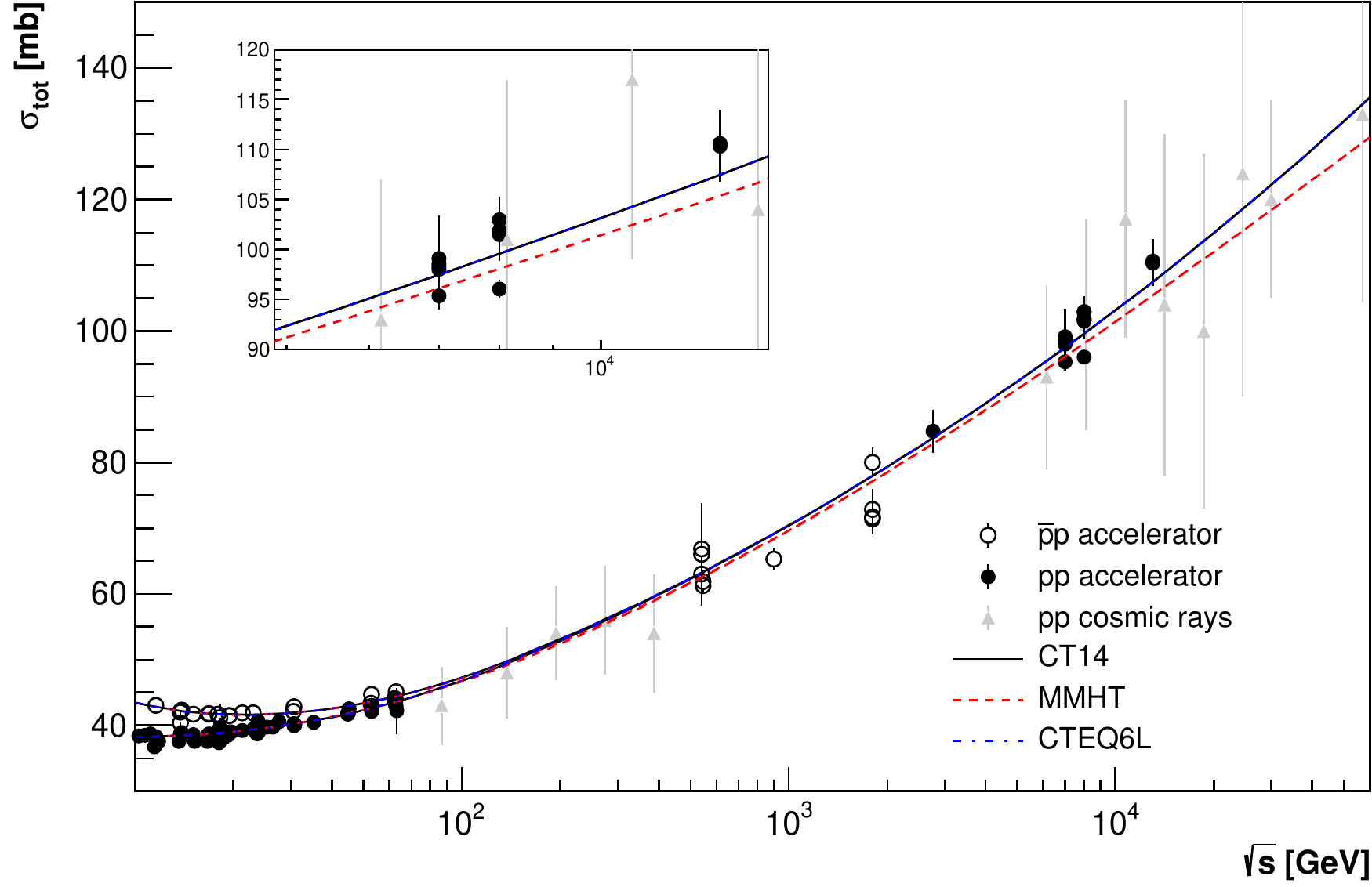}
\includegraphics*[width=8.5cm,height=8cm]{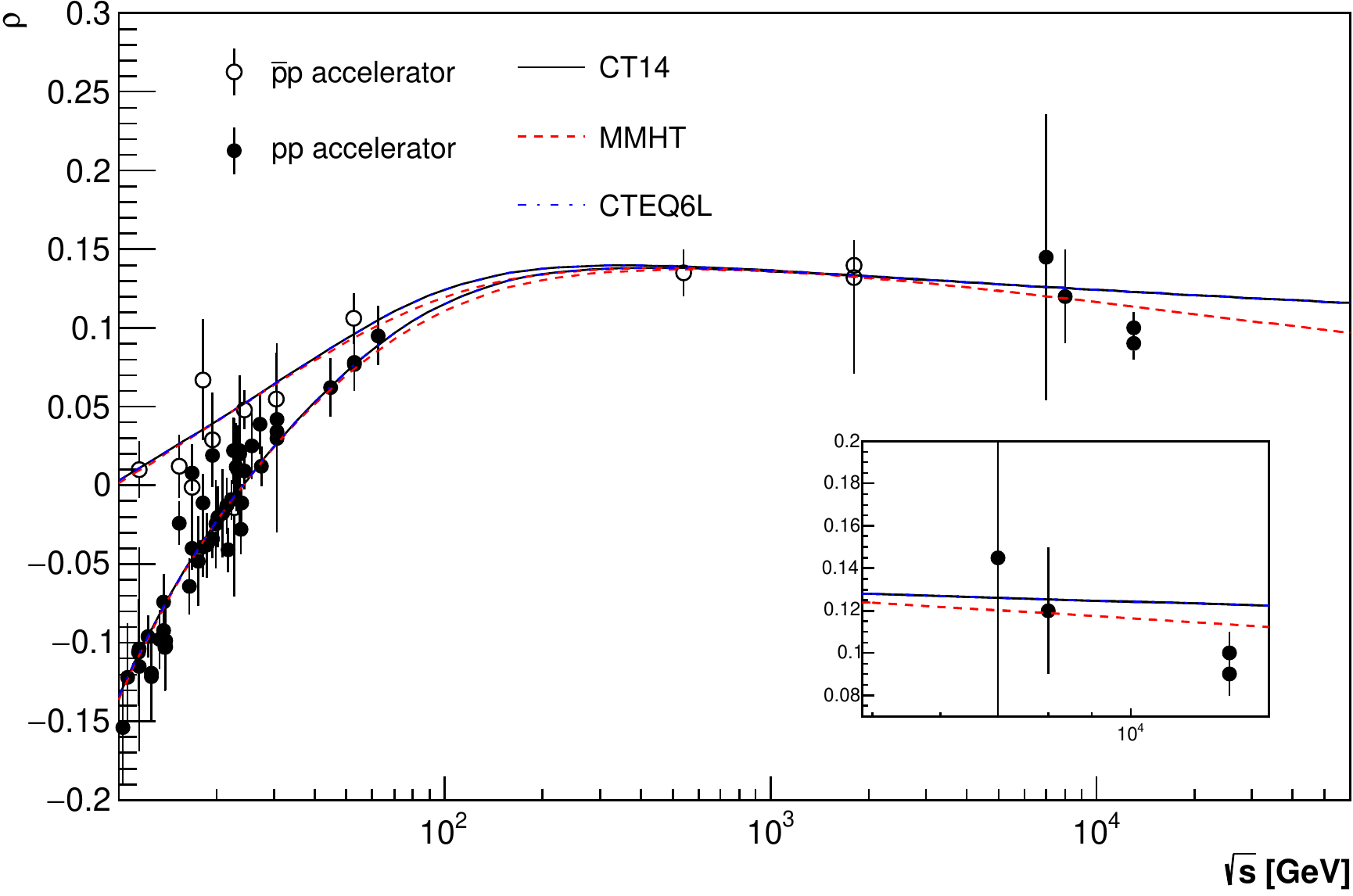}
\caption{Global 1$\sigma$-fit to $\sigma_{tot}^{pp/\bar{p}p}$ and $\rho^{pp/\bar{p}p}$ data without the ATLAS measurements, for low-energy cutoff $\sqrt{s_{min}}=10$ GeV.}
\label{fig:noATLAS_cuts10-13}
\end{figure}

\begin{figure}[h!]
\centering
\includegraphics*[width=8.5cm,height=8cm]{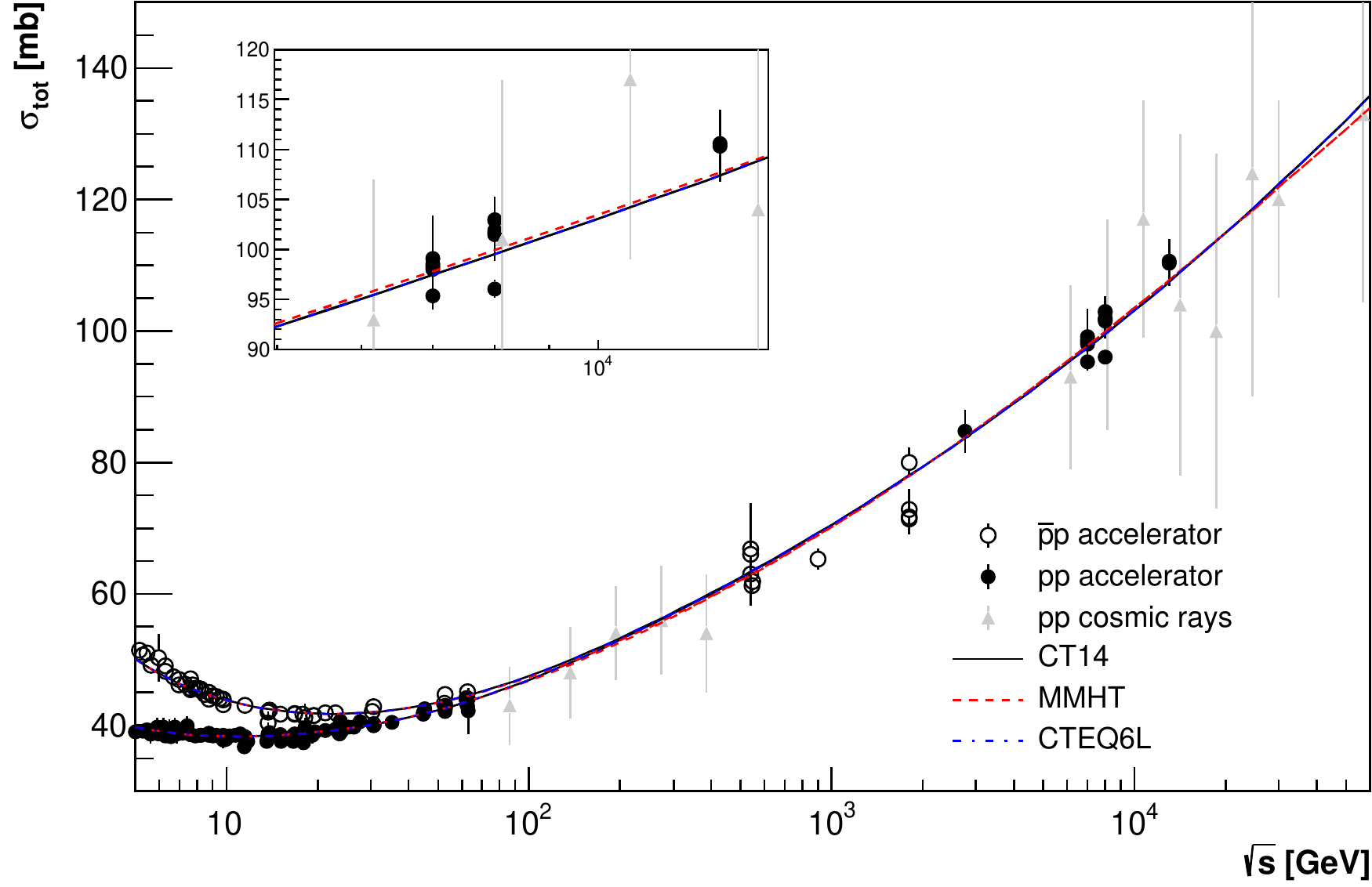}
\includegraphics*[width=8.5cm,height=8cm]{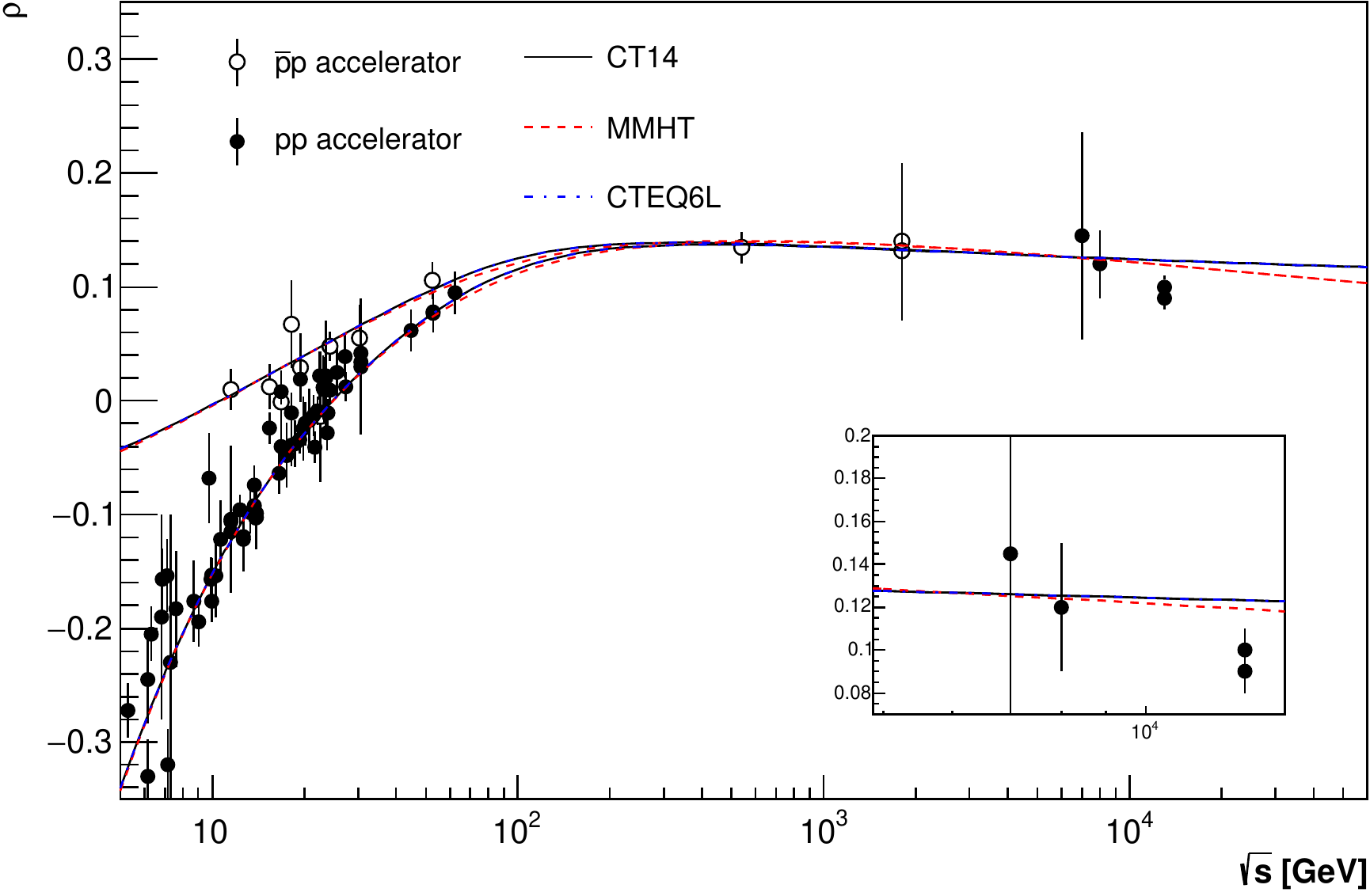}
\caption{Global 1$\sigma$-fit to $\sigma_{tot}^{pp/\bar{p}p}$ and $\rho^{pp/\bar{p}p}$ data without the ATLAS measurements, for maximum-energy cutoff $\sqrt{s_{max}}=13$ TeV and low-energy cutoff $\sqrt{s_{min}}=5$ GeV.}
\label{fig:noATLAS_cuts5-13}
\end{figure}

It is well known the discrepancies between the TOTEM and ATLAS data on $\sigma_{tot}$
at 7 and 8 TeV \cite{Fagundes:2017cmp}. Here we present two tests with low-energy cutoffs at
10 GeV and 5 GeV, in which the ATLAS data
are not included in the data set. The results are presented in Table \ref{tab:smin10GeV_semATLAS},  Figure \ref{fig:noATLAS_cuts10-13}
     ($\sqrt{s_{min}}$ = 10 GeV) and Table \ref{tab:smin5GeV_semATLAS}, Figure \ref{fig:noATLAS_cuts5-13}
     ($\sqrt{s_{min}}$ = 5 GeV).
     
     \begin{table}[h!]
%\scalebox{0.9}{
\centering
\caption{Best-fit parameters of our model without ATLAS data and low-energy cutoff $\sqrt{s_{min}}$ = 10 GeV. The number of degrees of freedom is $\zeta$ = 165.}
\vspace*{.2cm}
\begin{tabular}{c@{\quad}c@{\quad}c@{\quad}c@{\quad}}
\hline \hline
& & &  \\[-0.3cm]
PDF:    & CTEQ6L & CT14 & MMHT \\[0.05ex]
\hline
& & &   \\[-0.3cm]
$\mu^{+}_{soft}$ [GeV]& 0.90\,$\pm$\,0.18    & 0.90\,$\pm$\,0.17     & 0.90\,$\pm$\,0.20    \\[0.15cm]
$A$ [GeV$^{-2}$]      & 111\,$\pm$\,11    & 88.6\,$\pm$\,9.4     & 94\,$\pm$\,11    
\\[0.15cm]
$B$ [GeV$^{-2}$]      & 48\,$\pm$\,10    & 51.3\,$\pm$\,9.6     & 54\,$\pm$\,10     
\\[0.15cm]
$C$ [GeV$^{-2}$]      & 15.8\,$\pm$\,6.8    & 27.0\,$\pm$\,5.6    & 19.4\,$\pm$\,6.6    \\[0.15cm]
$\mu^{-}_{soft}$ [GeV]& 0.5 (fixed)          & 0.5 (fixed)           & 0.5 (fixed)          \\[0.15cm]
$D$ [GeV$^{-2}$]      & 24.2\,$\pm$\,1.4     & 24.2\,$\pm$\,1.4      & 24.2\,$\pm$\,1.4    \\[0.15cm]
$\nu_{1}$ [GeV]       & 1.61\,$\pm$\,0.20    & 1.69\,$\pm$\,0.20     & 1.44\,$\pm$\,0.20    \\[0.15cm]
$\nu_{2}$ [GeV]       & 0.0076\,$\pm$\,0.013  & 0.017\,$\pm$\,0.013   & -0.009\,$\pm$\,0.012  \\[0.15cm]
& & &   \\[-0.3cm]
\hline
& & &   \\[-0.3cm]
$\chi^2/\zeta$   & 1.278                  & 1.261                   & 1.251                    \\[0.15cm]
$P(\chi^2;\zeta)$    & 9.3 $\times$ 10$^{-3}$ & 1.3 $\times$ 10$^{-2}$ & 1.6 $\times$ 10$^{-2}$   \\[0.05ex]
& & &   \\[-0.3cm]
\hline \hline 
\label{tab:smin10GeV_semATLAS}
\end{tabular}
\end{table}

Comparing the results of Table \ref{tab:smin10GeV_semATLAS} and Fig. \ref{fig:noATLAS_cuts10-13} with the ones in Table \ref{t:paramsDGM} and Fig. \ref{fig:DGM_cut13} (with complete dataset) we see that, without the ATLAS data, the integrated probability increases as a consequence of the aforementioned discrepancies. On the one hand, it is interesting to note that the exclusion of the ATLAS data leads to a better description of $\sigma_{tot}$ at LHC13, both in visual and statistical grounds, for PDFs CT14 and CTEQ6L. On the other, MMHT prediction gives a lower cross section at 13 TeV, while providing a smaller $\rho$. Such behavior can be understood in the light of PDFs small-x extrapolation, as we discuss in the following.

A final interesting remark about these results lies in the fact that, by lowering the energy cutoff $\sqrt{s_{min}}$ to 5.0 GeV improves the description $\sigma_{tot}$ at LHC13 for all PDFs, as we note that in such case, the uncertainties in MMHT low-energy parameters ($A$ and $B$) are reduced. Yet, incompatibility of the model results with TOTEM's measurements of $\sigma_{tot}$ and $\rho$ at LHC13 are evident in these cases.

\begin{table}[h!]
%\scalebox{0.9}{
\centering
\caption{Best-fit parameters of our model without ATLAS data and low-energy cutoff $\sqrt{s_{min}}$ = 5.0  GeV. The number of degrees of freedom is $\zeta$ = 249.}
\vspace*{.2cm}
\begin{tabular}{c@{\quad}c@{\quad}c@{\quad}c@{\quad}}
\hline \hline
& & &  \\[-0.3cm]
PDF:    & CTEQ6L & CT14 & MMHT \\[0.05ex]
\hline
& & &   \\[-0.3cm]
$\mu^{+}_{soft}$ [GeV]& 0.90\,$\pm$\,0.15& 0.90\,$\pm$\,0.14     & 0.90\,$\pm$\,0.16    \\[0.15cm]
$A$ [GeV$^{-2}$]      & 92.9\,$\pm$\,6.8   & 82.3\,$\pm$\,6.0     & 88.4\,$\pm$\,6.8    \\[0.15cm]
$B$ [GeV$^{-2}$]      & 58.1\,$\pm$\,4.5     & 59.8\,$\pm$\,4.2      & 61.9\,$\pm$\,4.4     \\[0.15cm]
$C$ [GeV$^{-2}$]      & 20.9\,$\pm$\,4.5     & 30.5\,$\pm$\,3.8   & 22.6\,$\pm$\,4.6    \\[0.15cm]
$\mu^{-}_{soft}$ [GeV]& 0.5 (fixed)          & 0.5 (fixed)           & 0.5 (fixed)          \\[0.15cm]
$D$ [GeV$^{-2}$]      & 26.03\,$\pm$\,0.74   & 26.02\,$\pm$\,0.73    & 26.02\,$\pm$\,0.74   \\[0.15cm]
$\nu_{1}$ [GeV]       & 1.70\,$\pm$\,0.19    & 1.75\,$\pm$\,0.20     & 1.50\,$\pm$\,0.20    \\[0.15cm]
$\nu_{2}$ [GeV]       & 0.014\,$\pm$\,0.012  & 0.019\,$\pm$\,0.013   & -0.004\,$\pm$\,0.013   \\[0.15cm]
& & &   \\[-0.3cm]
\hline
& & &   \\[-0.3cm]
$\chi^2/\zeta$   & 1.422                  & 1.438                 & 1.395                    \\[0.15cm]
$P(\chi^2;\zeta)$    & 1.3 $\times$ 10$^{-5}$ & 7.0 $\times$ 10$^{-6}$ & 3.7 $\times$ 10$^{-5}$   \\[0.05ex]
& & &   \\[-0.3cm]
\hline \hline 
\label{tab:smin5GeV_semATLAS}
\end{tabular}
\end{table}

% \begin{figure}[H]

\section{Conclusions}

In this paper we have discussed recent studies on forward $pp$ and
$\bar{p}p$ elastic scattering within an eikonal QCD-based formalism. The model combines the
perturbative parton-model approach (to model the semihard interactions among partons), with a
Regge-inspired model (to describe the underlying soft interactions) and brings up information about the infrared properties of QCD by considering the possibility that the nonperturbative dynamics of QCD generate an effective charge.

We have presented a phenomenological analysis undertaken to improve the
understanding of elastic
processes taking place in the LHC. We address this issue by means of a
model involving only
even-under-crossing amplitudes at very high energies. As a result, we see
that the QCD-based model
allows us to describe the forward scattering quantities $\sigma_{tot}$ and
$\rho$ from
$\sqrt{s}=10$ GeV to 8 TeV in a quite satisfactory way, but not the TOTEM
measurements at 13 TeV
simultaneously.

Our analysis, which follows a previous short letter
\cite{Broilo:2019yuo}, explores in detail the various effects that could be important in the global fits, in special the use of three different PDFs (CT14,CTEQ6L and MMHT), investigating not only the difference and similarities among them, but also the effect of being pre or post LHC distributions.

On general grounds, from a statistical viewpoint, the present results demonstrate an overall satisfactory agreement of all PDFs with $\sigma_{tot}$ and
$\rho$ data over a wide range of energies. 
However, specifically at $\sqrt{s}=13$ TeV, our results for
$\rho$ ($\sigma_{tot}$) are greater (lower)
than the TOTEM measurements.
We understand that the inclusion of a crossing-odd elastic term in the scattering amplitude may improve the description of the
forward data at high energies. Such a result might be an indication that an Odderon does indeed have an important role in the soft and/or semihard interactions
at LHC energies\footnote{It is worth noting that some recent well known phenomenological approaches, with even-under-crossing dominance and without a satisfactory simultaneous
description of the TOTEM data at 13 TeV, indicate a different interpretation. For example,
Donnachie and Landshoff, within the Regge approach,  obtain for rho at 13 TeV the value 0.14
and conclude that  "there is no strong case for the presence of an odderon contribution to forward scattering" \cite{Donnachie:2019ciz}. Durand and Ha in the context of a QCD-based model, employ the sieve process proposed by Martin Block \cite{Block:2005qm}, which excludes the rho measurements at 13 TeV from the dataset as outliers \cite{Durand:2018irx}.}. We are presently investigating the subject.

\section*{Acknowledgments}

We thank Victor Gon{\c c}alves for useful discussions. This research was partially supported by the Conselho Nacional de Desenvolvimento Cient\'{\i}fico e Tecnol\'ogico (CNPq) under the grants 141496/2015-0 and 155628/2018-6,
by the project INCT-FNA Proc. No. 464898/2014-5, by the PEDECIBA program, and by the ANII-FCE-126412 project.

\appendix

\section{Parametrization for $\sigma_{QCD}(s)$}

One of the most important ingredient of the QCD-based model is the even-under-crossing 
partonic cross-section $\sigma_{QCD}(s)$, given by Eq. (\ref{sigQCD}). Here we
present the details  of the evaluation of this quantity,
using PDFs: CTEQ6L , CT14 and MMHT. Some additional results are also presented and discussed. 
The evaluation is based on the steps that follow.

First we consider the  complex analytic parametrization
\begin{eqnarray}
\sigma_{QCD}(s) &=& b_{1}+b_{2}\,e^{b_{3}[X(s)]^{1.01\, b_{4}}} \nonumber \\
&+& b_{5}\,e^{b_{6}[X(s)]^{1.05\, b_{7}}}+b_{8}\,e^{b_{9}[X(s)]^{1.09\, b_{10}}},
\label{eq:sigQCD_fit}
\end{eqnarray}
where $b_{1},...,b_{10}$ are free fit parameters and
\begin{equation}
X(s)=\ln\,\ln(-is)
\label{eq:X}
\end{equation}
provides the adequate complex and even character of the analytic function
through the substitution $s \rightarrow -is$, leading to $\mathrm{Re}\,\sigma_{QCD}(s)$
and $\mathrm{Im}\,\sigma_{QCD}(s)$.

Next,  by means of Eq. (\ref{sigQCD}) and using the three distinct PDFs,   
we generate around 30 points for each one of these parton distributions,
which are then fitted by the $\mathrm{Re}\,\sigma_{QCD}(s)$, with less than 1$\%$ error.
With the values of the free fit parameters determined for each PDF, the corresponding $\mathrm{Im}\,\sigma_{QCD}(s)$ are
evaluated.

\begin{table}[h!]
\label{tab:sig_QCD}
\centering
\caption{Fit results of the Re $\sigma_{QCD}$ in Eqs (1) and (2) to the actual data (see text).}
\begin{tabular}{c@{\quad}c@{\quad}c@{\quad}c@{\quad}}
\hline \hline
& & &  \\[-0.3cm]
PDF:    & CTEQ6L & CT14 & MMHT  \\[0.05ex]
\hline
& & & \\[-0.3cm]
$b_{1}$ [GeV$^{-2}$]  & 97.005                    & 100.220                  & 95.284                     \\[0.15cm]
$b_{2}$ [GeV$^{-2}$]  & 0.280 $\times$ 10$^{-1}$  & 0.434 $\times$ 10$^{-1}$ & 0.372                      \\[0.15cm]
$b_{3}$  & 1.699                     & 1.274                    & 0.600                      \\[0.15cm]
$b_{4}$  & 1.736                     & 1.919                    & 2.496                      \\[0.15cm]
$b_{5}$ [GeV$^{-2}$] & -0.149 $\times$ 10$^{-5}$ & 0.122 $\times$ 10$^{-7}$ & -0.255 $\times$ 10$^{-5}$  \\[0.15cm]
$b_{6}$  & 14.140                    & 14.050                   & 14.281                     \\[0.15cm]
$b_{7}$  & 0.319                     & 0.504                    & 0.281                      \\[0.15cm]
$b_{8}$ [GeV$^{-2}$] & 0.836 $\times$ 10$^{-1}$  & 3.699 $\times$ 10$^{3}$  & 0.909                      \\[0.15cm]
$b_{9}$  & 3.813                     & -80.280                  & 4.290                      \\[0.15cm]
$b_{10}$ & 0.810                     & -2.632                   & 0.673                      \\[0.15cm]
& & &  \\[-0.5cm]
\hline \hline 
\end{tabular}
%}
\label{tab:sigQCD_fit}
\end{table} 

For CTEQ6L, CT14 and MMHT we display in Table \ref{tab:sigQCD_fit} the best-fit parameters $b_{i}$, $i=1,\cdots, 10$  and 
in Fig. 9 the dependencies of $\mathrm{Re}\, \sigma_{QCD}(s)$ and $\mathrm{Im}\, \sigma_{QCD}(s)$.

From the figure, we see in all cases the steep rise of the partonic cross-sections with the energy.
For example at $\sqrt{s} = 10$ TeV, most results lie around 580 mb. Notice, however,
that this rise is tamed in the physical cross-sections, since we have an eikonalized model.
\vspace*{.3cm}
\begin{figure}[H]
\centering
\includegraphics*[width=8.5cm,height=8cm]{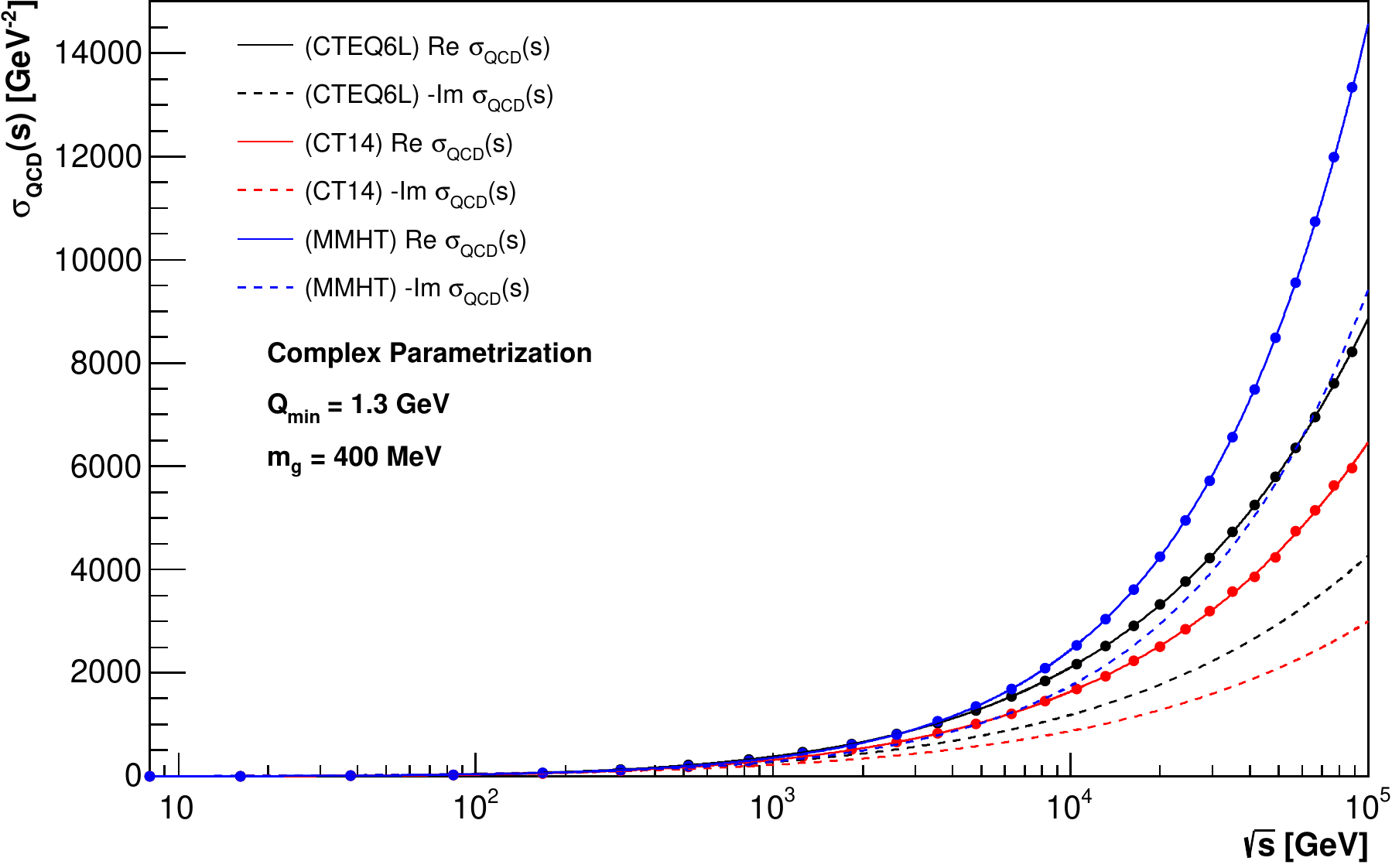}
\caption{ Real and imaginary parts of the complex $\sigma_{QCD}$ for each PDF. Small dots represent theoretical LO calculations from Eq. (\ref{sigQCD}) - roughly 30 points for each PDF. Solid curves correspond to Re $\sigma_{QCD}(s)$ fit,
Eqs. (A1) and (A2), to the data, represent by the dots, with less than 5$\%$ per datum $j=1,\cdots,30$. Dashed curves give Im $\sigma_{QCD}(s)$, as calculated from Eqs. (A1) and (A2), using fit parameters furnished in Table \ref{tab:sigQCD_fit}.}
\label{fig:sig_QCD}
\end{figure}

We note that among the PDFs post-LHC, MMHT led to the fastest rise of both
$\mathrm{Re}\, \sigma_{QCD}(s)$ and $\mathrm{Im}\, \sigma_{QCD}(s)$
and CT14 led to the slowest rise. The results with CTEC6L (pre-LHC) lie between these
two cases.

The extreme fast rise of $\sigma_{QCD}(s)$ in case of MMHT, may be the responsible
may be the responsible for the decrease of $\rho$ at
the LHC region (see Figure \ref{fig:DGM_cut13}).

\bibliographystyle{h-physrev}
\bibliography{paperBFLM_V3arXiv}

\end{document}